\documentclass[showpacs,preprintnumbers,amsmath,amssymb]{revtex4}
\usepackage[dvips]{graphicx}
\usepackage{dcolumn}
\usepackage{bm}

\begin{document}

\title
{Stability and Response of Vortex Solid Formed in Second Landau Level}

\author{Yuto Yokota, Dai Nakashima, and Ryusuke Ikeda}

\affiliation{Department of Physics, Graduate School of Science, Kyoto University, Kyoto 606-8502, Japan
}

\date{\today}

\begin{abstract} 
Physical properties of the vortex solid phase formed in the second Landau level (2LL), which may be stabilized by strong paramagnetic pairbreaking (PPB), are examined in type II limit with no magnetic screening. First, it is shown that the spectrum of the low energy mode of this vortex solid has the same form as that of the conventional vortex solid in the first (i.e., the lowest) Landau level. Using this result, the  melting line of the 2LL vortex solid is examined according to the Lindemann criterion. In contrast to the properties in equlibrium, the electromagnetic response of this vortex solid is quite unusual: Reflecting the presence of antivortices supporting the stability of the lattice structure, the superfluid stiffness measuring the response for a current perpendicular to the magnetic field is found to be nonvanishing, and its sign depends upon the applied current direction. Consequences of this response property are briefly 
discussed. 
\end{abstract}


\maketitle

\section{Introduction}
Recently, the presence of a high field and low temperature superconducting (SC) phase in the quasi two-dimensional iron-based superconductor FeSe has been argued based on thermal conductivity measurements in the field configuration parallel to the basal-plane \cite{Kasa20}. It is natural to expect, based on the strong paramagnetic pair-breaking (PPB) effect in this material, that this phase be a kind of Fulde-Ferrell-Larkin-Ovchinnikov (FFLO) phase \cite{FF,LO}. A remarkable feature seen upon entering this novel phase by sweeping the magnetic field is a negative magnetoresistance appearing only in the SC fluctuation regime. It has been clarified in Ref.\cite{NNI} that such a negative fluctuation magnetoresistance in the high field and low temperature region is an evidence of the SC fluctuation formed in the second Landau level (2LL) of the SC order parameter. It strongly suggests that the detected novel SC phase \cite{Kasa20} in the real sample of FeSe is the vortex solid or glass phase formed in 2LL of the SC order parameter. Although the spatial pattern of the amplitude of the SC order parameter in this 2LL vortex solid has been studied in previous works \cite{Klein,YM}, to the best of our knowledge, no theoretical results on physical properties of this state have been reported so far. 

In this work, the structures and physical properties of the 2LL vortex solid stabilized by moderately strong paramegnetic pair-breaking (PPB) are examined in a manner of comparing with those of the conventional vortex solid (lattice) formed in the first Landau level (1LL) \cite{AAA} which is the lowest LL in the usual situation with a negligibly small PPB. First, we point out that, through the spatial pattern of the gradient of the gauge-invariant phase obtained consistently with that of the amplitude \cite{Klein,YM}, the 2LL vortex solid has a highly anisotropic structure with a two-fold symmetry arising from mixing of {\it antivortices} with the winding number of the opposite sign to that of the field-induced vortices. To verify whether the 2LL vortex solid is stabilized in spite of coexistence with the antivortices or not, the harmonic fluctuation around the mean field solution of the 2LL vortex solid is examined in type II limit where the fluctuation of the flux density is negligible, and its dispersion relation is found to be, like that of the 1LL vortex solid, of ${\bf k}_\perp^4$-form with respect to the perpendicular component ${\bf k}_\perp$ to the magnetic field of the wave vector. Consequently, the phase correlation in the three-dimensional (3D) 2LL vortex solid in type II limit is quasi long-ranged \cite{Moore92,IOT92}, and the presence of a well-defined vortex liquid regime due to melting of the 2LL vortex solid \cite{IOT92} is expected. By examining the melting line according to the Lindemann criterion \cite{Houghton,Moore89,Nakashima}, a typical field v.s. temperature phase diagram of a 3D superconductor with strong PPB in higher fields is proposed. 

It is well accepted that, in a perfectly clean superconductor, a single vortex excitation flows under a homogeneous current. In addition, it is believed that, in the conventional $s$-wave superconductor, the unpinned and ordered triangular vortex lattice flows while keeping its structure unchanged under a homogeneous current, like in the case of a single vortex. However, it is quite unclear whether any ordered vortex solid shows a similar vortex flow motion. In fact, a single vortex flow can be regarded as a simplified picture on the vortex flow in the vortex {\it liquid}, while the ordered vortex solid is realized only through the freezing phase transition of the the disordered vortex liquid. 

In contrast to the properties in equilibrium which are qualitatively similar to those in the case of the vortex states in 1LL, the response properties of the 2LL vortex solid are found to be quite unconventional by examining the superfluid stiffness or the helicity modulus 
\begin{equation}
\Upsilon_{s} = \frac{\delta^2 F(\delta \bm{A})}{\delta A_i^2}, 
\label{ros}
\end{equation}
for a current flowing along the $i$-direction, where $F(\delta \bm{A})$ is the free energy functional under a disturbance $\delta \bm{A}$ of the electromagnetic gauge field $\bm{A}$. 
We find that, at the level of the linear response, the conventional vortex flow never occurs in the 2LL vortex solid as a result of the spontaneous anisotropy resulting from the mixing of the antivortices. Instead, depending on the relative direction of the applied current in the plane perpendicular to the magnetic field to the vortex lattice structure, the sign of $\Upsilon_s$ is changed. It implies that, depending on the current direction, the 2LL vortex solid is immobile so that the zero resistance (SC response) is supported, or the vortices flow with a structure instability accompanied. 

The present paper is organized as follows. In sec.II and III, derivation of the low energy mode and the vanishing superfluid stiffness in the conventional 1LL vortex solid is reviewed. In sec.IV, equilibrium properties of 2LL vortex lattice are studied, and the nonvanishing superfluid stiffness in 2LL vortex solid is explained in sec.V. In sec.VI, consequences of the nonvanishing superfluid stiffness in 2LL vortex solid are discussed, and some details on theoretical analysis are given in Appendices A and B. 

\section{Model and Equilibrium properties of 1LL vortex lattice}

First, key properties associated with the ordering of the conventional vortex lattice formed in the first LL (1LL) are reviewed in this and the next sections. To discuss theoretical results in both cases with moderately strong PPB and with a negligibly small PPB comprehensively, 
the following extended Ginzburg-Landau (GL) model \cite{YM} will be used throughout the main text of this paper : 
\begin{widetext}
\begin{equation}
{\cal H} = \int d^3\bm{r} \left[ c_1 \xi_0^2 |\bm{\Pi} \Delta|^2 - c_2 \xi_0^4 |\bm{\Pi^2} \Delta|^2 - \varepsilon_0 |\Delta|^2 
+ \frac{g}{2} |\Delta|^4 + \frac{g_1}{2} \xi_0^2 |\Delta|^2 |\bm{\Pi} \Delta|^2 + \frac{g_2}{2} \xi_0^2 ( \, (\Delta^*)^2 (\bm{\Pi} \Delta)^2 + {\rm c.c.} \, ) \right], 
\label{GL1}
\end{equation}
\end{widetext}
where $\Delta$ is the SC order parameter, $\xi_0$ is the GL coherence length, $\bm{\Pi} = - i \bm{\nabla} + 2 \pi \bm{A}/\phi_0$ is the gauge-invariant gradient operator, $\phi_0 = \pi c \hbar/|e|$ is the flux quantum for the charge $2e$, and $\varepsilon_0 = {\rm ln}(T_c(0)/T)$ which is always positive in any situation of our interest below. 


Throughout this work, we focus on the situations under an external magnetic field parallel to the $z$-axis so that the Landau gauge $\bm{A}_{\rm ext} = B y {\hat x}$ is used. Further, the difference $\bm{A} - \bm{A}_{\rm ext}$, denoted as $\delta \bm{A}$, is perpendicular to the applied uniform magnetic field and 
will be assumed to have been introduced externally. The dependences of the coefficients other than $\varepsilon_0$ on the temperature $T$ will be neglected. The eigenvalue of $\bm{\Pi}^2$ for any $(n+1)$-th LL state is $(2n+1) r_B^{-2}$, where $r_B = \sqrt{\phi_0/(2 \pi B)}$. 

In lower fields where the PPB effect is so small, the coefficient $c_1$ is positive, and consequently, 1LL is the lowest energy level of the order parameter field $\Delta$. Then, the mass (energy) gap between 1LL and 2LL 
\begin{equation}
r_{10} = 2 h (c_1 - 4 h c_2)
\end{equation}
may be assumed to be positive in the field range which is our focus in this and next sections, where $h= \xi_0^2/r_B^2 = 2 \pi \xi_0^2 B/\phi_0$. In fact, by constructing the GL Hamiltonian for the weak-coupling BCS model with PPB effects on the basis of the gradient expansion, one finds that not only $c_1$ but also $c_2$ are positive in low enough fields, and that the two coefficients $c_1$ and $c_2$ change their sign with increasing the field, i.e., as the PPB effect is enhanced. Concretely, within the weak-coupling BCS model, $c_2$ changes its sign with increasing $h$, and the sign of $c_1$ tend to be reversed at a higher field, while keeping $r_{10}$ positive. In the field range mentioned above, the depairing field $H_{c2}(T)$ is determined by 1LL modes of $\Delta$. The situation in which both $c_1$ and $c_2$ are negative due to strong PPB effect will be considered in sec.IV and V. 

It is well known that, in the presence of a negligibly weak PPB effect, the vortex lattice near $H_{c2}$ is expressed by one of basis functions $\varphi_0({\bf r}|{\bf r}_0)$ in 1LL of the order parameter $\Delta$. Below, this mean field vortex lattice solution in 1LL will be represented as 
\begin{equation}
\Delta_{0} = A_0 \varphi_0(\bm{r}|0), 
\label{Abrikosov0}
\end{equation}
where $\varphi_0(\bm{r}|\bm{r}_0) = e^{i y_0 {\overline x}/r_B} \varphi_0(\bm{r}+\bm{r}_0|0)$ \cite{Eilenberger}, and 
\begin{equation}
\varphi_0(\bm{r}|0) = e^{-{\overline y}^2/2} \sigma_0({\overline x}+i{\overline y}) 
\label{Abrikosov}
\end{equation}
with 
\begin{equation}
\sigma_0(\zeta) = \left(\frac{k^2}{\pi} \right)^{1/4} \sum_l \, \exp\left( - \frac{k^2 l^2}{2} + i \pi R \, l^2 \right) \exp(i k l \zeta). 
\end{equation}
Here $l$ takes interger values, $2 \pi r_B/k$ is the period in the $x$-direction, and $R$ is a parameter determining shape of the parallelogram expressing the unit cell (see Fig.1). Since the length scale was normalized by $r_B$ so that ${\overline {\bf r}} =$ (${\overline x}$, ${\overline y}$) $= {\overline {\bf r}} = {\bf r}/r_B$, $k$ as well as $R$ is dimensionless. 

\begin{figure}[t]
\scalebox{0.3}[0.3]{\includegraphics{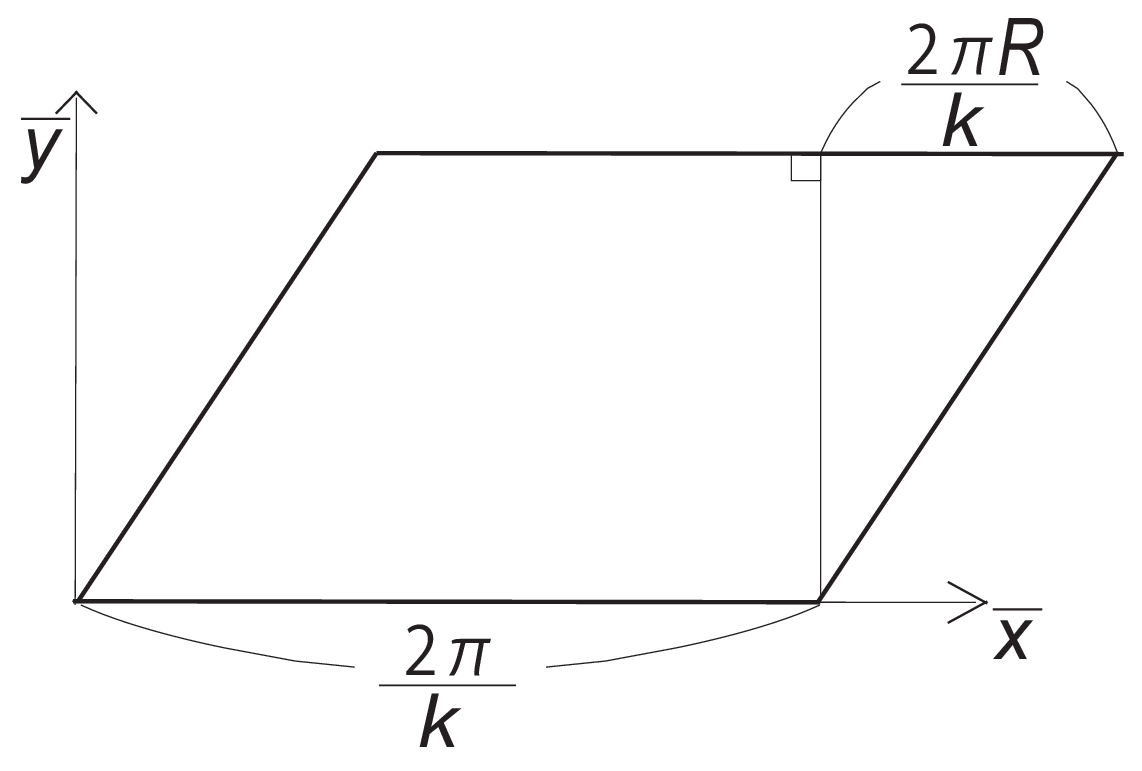}}
\caption{Parallelogram expressing the unit cell of the vortex lattice represented by $\varphi_0(\bm{r}|0)$. In unit of $r_B^2$, its area is $2 \pi$ due to flux quantization, and the parameters $k$ and $R$ are defined through this figure. }
\label{fig.1}
\end{figure}

The fact that $\sigma_0$ in eq.(\ref{Abrikosov}) is independent of $\zeta^* = {\overline x} - i {\overline y}$ implies that, for any 1LL vortex state expressed by a linear combination of $\varphi_0(\bm{r}|\bm{r}_0)$, all nodes in the $x$-$y$ plane express field-induced vortices, and that antivortices are not present in any 1LL state. 

For simplicity, we work in type II limit, where the magnetic screening in the mean field solution is neglected. Further, $\delta \bm{A}$ appearing in the ensuing description always implies an external disturbance of the electromagnetic gauge field. 

The function (\ref{Abrikosov}) is already normalized as $\langle |\varphi_0(\bm{r}|0)|^2 \rangle_s = 1$, where $\langle \,\,\, \rangle_s$ denotes the space average. The mean field solution is obtained by minimizing the expression obtained by replacing $\Delta$ in ${\cal H}$ by $\Delta_0$ with respect to $|A_0|^2$, $k$, and $R$. Then, the mean field equation resulting from minimizing with respect to $|A_0|^2$ is 
\begin{equation}
- \varepsilon_{h,0} + |A_0|^2 \biggl(g + \frac{g_1 h}{2} \biggr) \langle 0, 0|0, 0 \rangle = 0. 
\label{MFEQ0}
\end{equation} 
Here, 
\begin{equation}
\varepsilon_{h,0} = \varepsilon_0 -c_1h+c_2 h^2,  
\end{equation}
\begin{equation}
\langle n, m|p, s \rangle \equiv \langle (\varphi_n(\bm{r}|0) \varphi_m(\bm{r}|0))^* \varphi_p(\bm{r}|0) \varphi_s(\bm{r}|0) \rangle_s , 
\label{bracketMF}
\end{equation}
and the relation 
\begin{equation}
\langle 1, 0|1,0 \rangle = \frac{\langle 0, 0|0, 0 \rangle}{2}
\label{GLMaki}
\end{equation}
was used. Derivation of eq.(\ref{GLMaki}) will be shown in Appendix A. 
In eq.(\ref{bracketMF}), $\varphi_n$ is the basis function corresponding to $\varphi_0$ in the ($n+1$)-th LL given by 
\begin{equation}
\varphi_n(\bm{r}|0) = \frac{1}{\sqrt{n!}} ({\hat a}^\dagger)^n \varphi_0(\bm{r}|0), 
\end{equation}
where ${\hat a}^\dagger$ is the operator raising the Landau level (see Appendix A). 

The combination $g + g_1 h/2$ is positive in the low field regime of our focus in this section, because the gradient contributions proportional to $g_1$ and $g_2$ of the quartic term of the GL Hamiltonian are smaller than the $g$-term there. Further, the $g_2$ term is ineffective as far as just the order parameter modes in 1LL are taken into account. Therefore, 
the structure to be realized in equilibrium is, as in the conventional GL model in which $c_2$, $g_1$, and $g_2$ terms are absent, the triangular vortex lattice minimizing the so-called Abrikosov factor $\beta_A = \langle 0, 0|0, 0 \rangle$, where 
\begin{equation}
\langle 0, 0|0, 0 \rangle = \frac{k}{\sqrt{2 \pi}} \sum_{n,m} e^{-k^2(n^2+m^2)/2} {\rm cos}(2 \pi R n m). 
\label{Abrikosovbeta}
\end{equation}
It is well known that the parameter values in the triangular lattice are $k = \pi^{1/2} 3^{1/4}$ and $R=1/2$. 

To examine the stability of the mean field solution (\ref{Abrikosov0}), the intra LL fluctuation \cite{Eilenberger} 
\begin{equation}
\delta \Delta_0 = A_0[ a_{+0} \varphi_0(\bm{r}|\bm{r}_0) +  a_{-0} \varphi_0(\bm{r}|- \bm{r}_0) ]
\end{equation}
will be introduced. The vector $\bm{r}_0$ introduced here plays the role of the perpendicular component $\bm{q}_\perp$ to the applied magnetic field of the wave vector of the fluctuation, and, in fact, $\bm{q}_\perp$ is given by $\bm{r}_0 \times {\hat z}/r_B^2$. The harmonic fluctuation contribution in the GL model (\ref{GL1}) due to $\delta \Delta_0$ is given by replacing $\Delta_{\rm MF}$ and $\delta \Delta$ in ${\cal H}_{\Delta,1}$ and ${\cal H}_{\Delta,2}$ of (\ref{harmgl}) in Appendix by $\Delta_{0}$ and $\delta \Delta_0$, respectively. We note that the intra LL fluctuation does not couple with the gauge field disturbance $\delta {\bm{A}}$. Consequently, we have the two eigen-modes with eigen values 
\begin{equation}
\xi_\pm({\bf r}_0) \equiv \frac{\varepsilon_{\pm,0}}{g |A_0|^4} =  2 \xi_d({\bf r}_0) - \xi_0 \pm |\xi_n({\bf r}_0)|, 
\end{equation}
where 
\begin{eqnarray}
2 \xi_d({\bf r}_0) - \xi_0 &=& 2 \langle 0+,0|0+,0 \rangle - \langle 0,0|0,0 \rangle  + \frac{g_1 h}{g} ( \, \langle 1+, 0|1+, 0 \rangle + ({\rm Re}\langle 1+, 0|0+, 1 \rangle) \,  - \langle 1, 0|1, 0 \rangle \, ) \nonumber \\ 
\xi_n({\bf r}_0) &=& \langle 0,0|0+,0- \rangle + \frac{g_1 h}{g} \langle 1, 0|1+, 0- \rangle. 
\label{shear0}
\end{eqnarray}
Here, appreviated notation on the brackets 
\begin{eqnarray}
\langle n, m|p+, s- \rangle &=& \langle (\varphi_n(\bm{r}|0) \varphi_m(\bm{r}|0))^* \varphi_p(\bm{r}|\bm{r}_0) \varphi_s(\bm{r}|-\bm{r}_0) 
\rangle_s, \nonumber \\
\langle n+, m|p+, s \rangle &=& \langle (\varphi_n(\bm{r}|\bm{r}_0) \varphi_m(\bm{r}|0))^* \varphi_p(\bm{r}|\bm{r}_0) \varphi_s(\bm{r}|0) \rangle_s. 
\end{eqnarray}
were introduced. Within 1LL, there is no contribution from the $g_2$ term in $\varepsilon_{\pm,0}$. 
The expression (\ref{shear0}) is a simple extention of that in the conventional case with $g_1=g_2=0$ \cite{Eilenberger,Moore89}. In paticular, the minus (massless) mode with $\varepsilon_{-,0}$ consists primarily of the phase fluctuation of $\Delta$ and physically expresses the elastic energy of the shear deformation of the triangular vortex lattice \cite{Moore89,RI90}. 

It is sufficient to focus on the minus mode $\varepsilon_{-,0}$ to examine the stability. By expanding $\varepsilon_{-,0}$ in powers of ${\bf r}_0$, the absence of O(${\bf r}_0^2$) terms is easily checked. In fact, up to O(${\bf r}_0^2$), $|\xi_n({\bf r}_0)|$ in (\ref{shear0}) becomes 
\begin{equation}
\biggl( 1 + \frac{g_1 h}{2g} \biggr)[\langle 0,0|0,0 \rangle (1 - {\bf q}_\perp^2 r_B^2) - {\rm Re}(\langle 1,1|0,0 \rangle q_+^2 r_B^2)]
\label{shearenergy0},
\end{equation} 
and, up to O(${\bf r}_0^2$), $\varepsilon_{-,0}$ itself becomes zero, where $q_\pm = q_x \pm i q_y$. Therefore, the energy arising from in-plane distortions of $\Delta$ of the minus (massless) mode is proportional to $\bm{q}_\perp^4$. As has been clarified through the two derivations \cite{Moore89,Labush} of the shear energy, the stability is also ensured through minimizing the mean field free energy with respect to the parameters $k$ and $R$. By incorporating the dependence of the fluctuation energy on the out-of-plane component $q_z$ of the wave vector, the fluctuation energy due to the minus mode in the 1LL triangular lattice is expressed in the case with no PPB effect by 
\begin{equation}
\delta {\cal H}_{0, -} = \sum_{\bf q} \biggl[ \, N(0) |A_0|^2 \xi_0^2 \, q_z^2 + \frac{1}{2} C_{66} {\bf q}_\perp^4 r_B^4 \biggr] |a_{-0}|^2. 
\label{minusflucenergy}
\end{equation}
The expression of the shear modulus $C_{66}$ in the conventional GL model with $g>0$, and $g_1=g_2=0$ is given by 
\begin{equation}
C_{66} = 2 g |A_0|^4 0.119. 
\label{shearconst}
\end{equation}
Then, since, in the 1LL vortex lattice, the Fourier transformation ${\bf s}^{\rm T}_{\bf q}$ of the transverse component of the displacement field is given \cite{Moore89,RI90} 
by $r_B^2 ( i {\bf q} \times a_{-0} {\hat z})$, the Lindemann criterion for detemining the vortex lattice melting line takes the form 
\begin{equation}
\langle |\nabla \times \chi({\bf r}) {\hat z}|^2 \rangle = c_{\rm L}^2 r_B^{-2}, 
\label{Lindemann0}
\end{equation}
where $a_{-0}$ was identified with the Fourier transformation of the phase fluctuation $\chi({\bf r})$, and the constant $c_{\rm L}$ is taken to be less than unity. 

Before ending this section, the brackets $\langle n,m|p,s \rangle$ appearing repeatedly in our analysis will be defined concretely. The brackets with $n+m=p+s \pm 2$ are called hereafter as {\it asymmetric brackets}. The bracket $\langle 1,1|0,0 \rangle$ appeared above is one of the asymmetric ones. On the other hand, the brackets with $n+m=p+s$ including the familiar Abrikosov factor $\beta_A= \langle 0,0|0,0 \rangle$ are called as {\it symmetric brackets}. 

Although, in the triangular lattice with hexagonal symmetry, the asymmetric bracket $\langle 1,1|0,0 \rangle$ is zero so that the last term of (\ref{shearenergy0}) is lost, this term was kept here in relation to discussion to be performed in the ensuing sections.

\section{Uniform Flow Response of Vortex Lattice in 1LL} 

It is believed that, in the ideal pinning-free system, the conventional vortex lattice shows the vortex flow response under a uniform current perpendicular to the magnetic field of the same type as that of a single vortex. Since our purpose is to examine the response of the vortex lattice to a uniform current, it is sufficient to focus on the case with a disturbance of the gauge field $\delta \bm{A}$ with vanishing wave vector and in a direction perpendicular to the magnetic field. In the conventional triangular lattice, each vortex forming the vortex lattice is in a common environment, and hence, it is naturally understood that the property of the electromagnetic response is the same as that for a single vortex. This vortex flow response is described by the Josephson relation \cite{Josephson} 
\begin{equation}
\delta \bm{A} = \bm{s} \times \bm{B}, 
\label{phaseslip}
\end{equation}
which is equivalent to the phase slippage equation in the context of the neutral superfluid. In fact, according to the relation $\bm{E} = \bm{B} \times \partial \bm{s}/\partial (ct)$ equivalent to eq.(\ref{phaseslip}), the vortex motion with nonzero velocity $\partial \bm{s}/\partial t$ implies the appearance of an electric field $\bm{E}$. 

Further, at the static level, the Josephson relation implies that the superfluid stiffness, i.e., the helicity modulus (\ref{ros}), for the disturbance of the gauge field applied in a direction perpendicular to $\bm{B}$ is zero. Let us review derivation \cite{IOT92,RI95,RI07} in the GL model 
of the vanishing of $\Upsilon_s$ for a current flowing 
in the $i$-direction ($i=x$, $y$) in the stable triangular vortex solid phase formed in 1LL. 
For this purpose, the harmonic change of ${\cal H}$ occurring when a disturbance $\delta \bm{A} = \bm{A} - \bm{A}_{\rm ext}$ is introduced will be studied to find the superfluid stiffness (\ref{ros}). Here, since we focus here only on the electromagnetic response of the mean field vortex lattice, we only have to take account of the fluctuation of $\Delta$ coupling to the gauge disturbance $\delta \bm{A}$. 
For the 1LL vortex lattice, such a fluctuation is the mode of $\Delta$ in 2LL, and hence, we introduce 
\begin{equation}
\delta \Delta_1 = A_0 \, a_1 \varphi_1(\bm{r}|0). 
\label{fluc1}
\end{equation}
Hereafter, the gauge disturbance will be expressed by the dimensionless quantity \begin{equation}
\delta {\tilde {\bm{A}}} = \frac{2 \pi}{\phi_0} \xi_0 \, \delta \bm{A}. 
\label{gaugefluct}
\end{equation}
By identifying (\ref{Abrikosov0}) and (\ref{fluc1}) with $\Delta$ and $\delta \Delta$ in eq.(\ref{harmgl}) in Appendix. respectively, we find 
\begin{widetext}
\begin{eqnarray}
{\cal H}_{\Delta,1} &=& |A_0|^2 \biggl[- \varepsilon_{h,0} + 2h(c_1 - 4 h c_2) + 2 g |A_0|^2 \langle 1,0|1,0 \rangle \biggr] |a_1|^2 + \frac{g}{2} |A_0|^4 \biggl( \langle 0,0|1,1 \rangle a_1^2 + {\rm {c.c.}} \biggr), \nonumber \\
{\cal H}_{\Delta,2} &=& \frac{h|A_0|^4}{2} \biggl[ \, g_1 \biggl((\langle 1,1|1,1 \rangle + \langle 0,0|0,0 \rangle + 2 \langle 2,0|2,0 \rangle + 2 \sqrt{2} \langle 2,0|1,1 \rangle)|a_1|^2 + \sqrt{2}(\langle 1,0|2,1 \rangle a_1^2 + {\rm {c.c.}}) \biggr) \nonumber \\
&+& g_2 \biggl( 8 \langle 1,0|1,0 \rangle |a_1|^2 + 2 \sqrt{2}(\langle 0,0|2,0 \rangle a_1^2 + {\rm {c.c.}}) \biggr) \biggr], \nonumber \\
{\cal H}_{A,1} &=& |A_0|^2 \biggl[ (c_1 - 4 h c_2) \delta {\tilde A}_+ \delta {\tilde A}_- + \sqrt{\frac{h}{2}} \biggl(\delta {\tilde A}_+ \, a_1^* (c_1 - 2 c_2 h + c_1 - 6 c_2 h) + {\rm {c.c.}} \biggr) \biggr], \nonumber \\
{\cal H}_{A,2} &=& |A_0|^4 \biggl(\frac{g_1}{2} + g_2 \biggr) \biggl[ \langle 0,0|0,0 \rangle \delta {\tilde A}_+ \delta {\tilde A}_- + \sqrt{\frac{h}{2}} \biggl(\delta {\tilde A}_+ a_1^* (2 \langle 1,0|1,0 \rangle + \langle 0,0|0,0 \rangle) + {\rm {c.c.}} \biggr) \biggr] \nonumber \\ 
&+& |A_0|^4 \sqrt{\frac{h}{2}} \biggl(\frac{g_1}{2} + g_2 \biggr) \biggl( \delta {\tilde A}_+ a_1 (\langle 0,0|1,1 \rangle + \sqrt{2} \langle 0,0|2,0 \rangle) + {\rm {c.c.}} \biggr) \nonumber \\
&=& \biggl(\frac{g_1}{2} + g_2 \biggr) |A_0|^4 \langle 0,0|0,0 \rangle \, [ \delta {\tilde A}_+ \delta {\tilde A}_- + \sqrt{2h} (\delta {\tilde A}_+ a_1^* + {\rm {c.c.}}) \, ], 
\label{fluctochu0}
\end{eqnarray}
\end{widetext}
where the relation $\langle 0,0|1,1 \rangle = - \sqrt{2} \langle 0,0|2,0 \rangle$ given in eq.(\ref{bra2}) in Appendix was used. By using eq.(\ref{MFEQ0}) and eqs.(\ref{bra1}) and (\ref{bra2}) in Appendix, the expression of ${\cal H}_{\Delta,1} + {\cal H}_{\Delta,2}$ is rewritten in the following form with no $\varepsilon_{h,0}$, i.e.,  
\begin{widetext}
\begin{equation}
{\cal H}_{\Delta,1} + {\cal H}_{\Delta,2} = |A_0|^2 \biggl( 2h(c_1 - 4 c_2 h) + (g_1+2 g_2) h |A_0|^2 \langle 0,0|0,0 \rangle \biggr)|a_1|^2 + \frac{\sqrt{2}}{4} |A_0|^4 (h(4 g_2 - g_1) - 2g) [\langle 0,0|2,0 \rangle a_1^2 
+ {\rm {c.c.}} ]. 
 \end{equation}

By summing up all the contributions examined above, the harmonic fluctuation contribution associated with the gauge disturbance $\delta \bm{A}$ to the GL Hamiltonian finally becomes 
\begin{equation}
\delta {\cal H} = 2 h |A_0|^2 \biggl[ (c_1 - 4 c_2 h) + \biggl(\frac{g_1}{2} + g_2 \biggr) |A_0|^2 \langle 0,0|0,0 \rangle \biggr] |{\tilde a}_1|^2 + \frac{|A_0|^4}{4}(2g+h(g_1 - 4 g_2)) \biggl[\langle 0,0|1,1 \rangle a_1^2 + {\rm {c.c.}} \biggr], 
\label{flowLLL}
\end{equation}
\end{widetext}
where 
\begin{equation}
{\tilde a}_1 = a_1 + \frac{\delta {\tilde A}_+}{\sqrt{2h}}. 
\end{equation}
If $a_1$ is identified with $(s_y - i s_x)/(\sqrt{2} \, r_B)$, the expression (\ref{phaseslip}) is nothing but the equality ${\tilde a}_1=0$ in the present notation. 

Except the presence of the last term in (\ref{flowLLL}), the free energy used in eq.(\ref{ros}) is independent of $\delta {\bf A}$ after integrating over $a_1$, implying that the equality ${\tilde a}_1 = 0$ is satisfied. Actually, the bracket $\langle 0,0|1,1 \rangle$ vanishes \cite{Lasher} in the six-fold symmetric triangular lattice and the four-fold symmetric square lattice, and thus, the relation (\ref{phaseslip}) is satisfied. In other words, $\Upsilon_s$ in a current perpendicular to the magnetic field is zero. In this manner, the conventional picture that the vortex flow response occurs in the case of the lattice structure minimizing the energy is justified. However, the bracket $\langle 0,0|1,1 \rangle$ does not vanish when the realized vortex lattice has a spontaneous anisotropy and is merely two-fold symmetric. In such a case where the last term in (\ref{flowLLL}) is nonvanishing, the relation (\ref{phaseslip}) is not satisfied. That is, it is anticipated that the conventional vortex flow response may not occur in a vortex lattice formed with a spontaneous anisotropy. 

\section{Equilibrium Properties of Vortex Lattice in 2LL}

Theoretical research works \cite{Klein,YM,AI03,Buzdin} performed so far have clarified that the vortex lattice to be realized in the presence of a strong PPB effect may become a solution of $\Delta$ in the second LL (2LL). In fact, it has been argued recently \cite{NNI} that the negative magnetoresistance appearing in the force-free configuration and only in the low temperature fluctuation regime of the iron-based superconductor FeSe \cite{Kasa20} is an evidence of the superconducting fluctuation belonging to 2LL. This peculiar fluctuation effect strongly suggests that the novel high field superconducting phase \cite{Kasa20} lying at low temperatures of this material is a vortex solid belonging to 2LL. Although the structure of this vortex lattice has been discussed previously \cite{Klein,YM,AI03}, to the best of our knowledge, results on physical properties in this phase have not been reported so far. 

The model (\ref{GL1}) will be used in higher fields where a remarkable PPB effect is present. As already mentioned, the coefficients $c_1$ and $c_2$ appearing in the model in this situation tend to have different signs compared with those in low enough fields. To focus on such a high field range in which the $H_{c2}(T)$ curve and the vortex lattice near $H_{c2}$ are determined by 2LL modes of $\Delta$, both of $c_1$ and $c_2$ are assumed to be negative in this section. For this reason, $c_j$ ($j=1$ or $2$) will be rewritten below as $-|c_j|$, and, for simplicity, their dependences on the temperature $T$ and the flux density $B$ will be neglected hereafter. 
Within the model (\ref{GL1}), the field range in which the $H_{c2}(T)$ curve is determined by 2LL states is given by 
\begin{equation}
\frac{|c_1|}{8 |c_2|} < h < \frac{|c_1|}{4 |c_2|}. 
\label{secondllregion}
\end{equation} 
As is seen below in various places, the "mass difference" $r_{01}$ between 1LL and 2LL, which is equivalent to the relative energy of 1LL measured from 2LL, and the corresponding one $r_{21}$ between the third LL (3LL) and 2LL are given by 
\begin{eqnarray}
r_{01} &=& 2h (|c_1| - 4 |c_2| h), \nonumber \\ 
r_{21} &=& 2h (8|c_2|h - |c_1|). 
\end{eqnarray}

\begin{widetext}
\begin{figure}[t]
\scalebox{0.7}[0.7]{\includegraphics{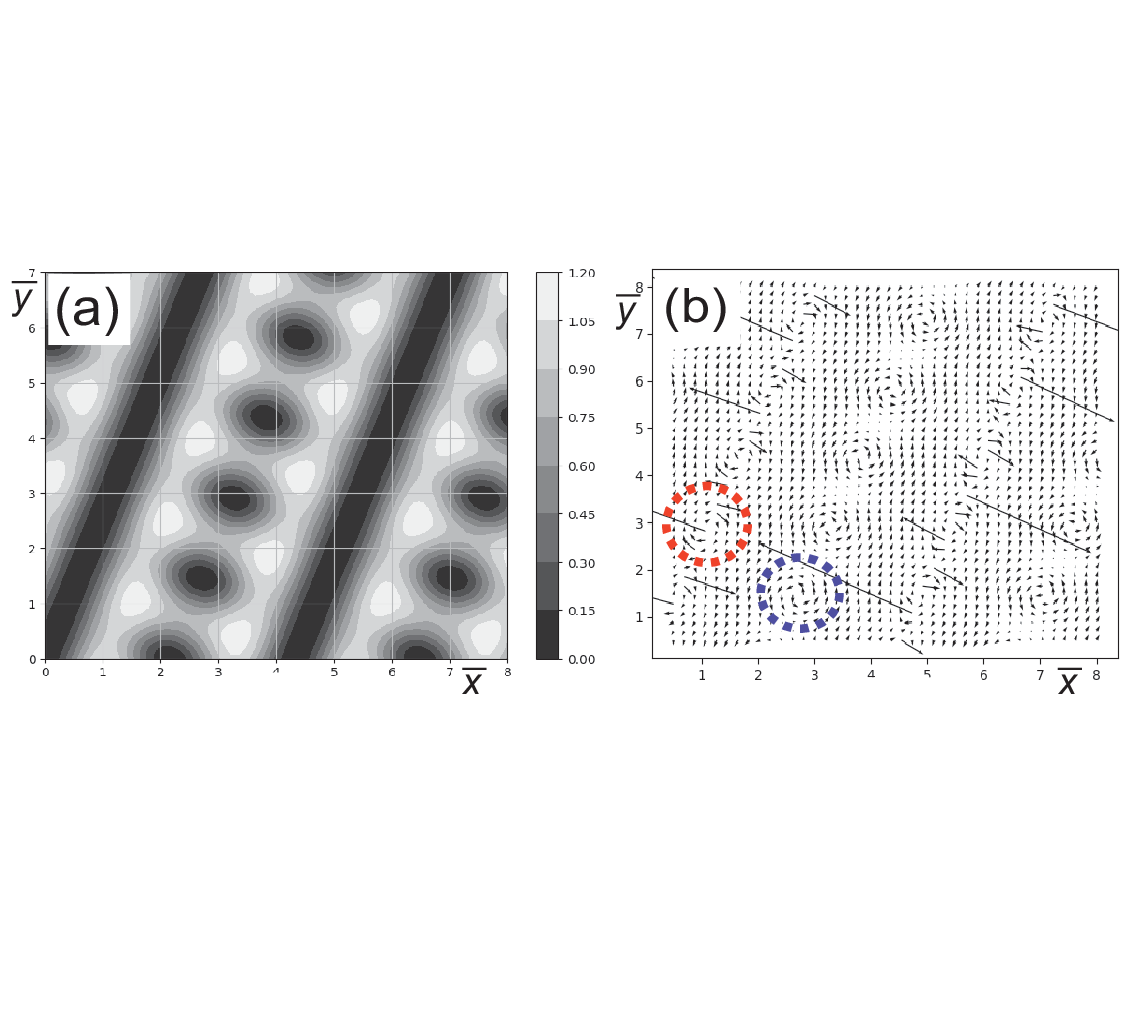}}
\caption{Spatial variations of (a) the amplitude $|\Delta_1|$ and (b) the gradient of the gauge-invariant phase, $({\rm Re} \Delta_1^* \bm{\Pi} \Delta_1)/|\Delta_1|^2$, obtained in terms of the parameters $g_1 h/g = -0.6$ and $g_2 h/g = 0.15$ with $g > 0$. The resulting parameters determining 2LL vortex lattice structure are $k=1.4535$ and $R = 0.12862$. One of the conventional field-induced vortices and an antivortex are indicated by the blue dashed circle and the red dashed one, respectively. }
\label{fig.2}
\end{figure}
\end{widetext}

First, let us start from determining the mean field solution 
\begin{equation}
\Delta_1 = A_1 \varphi_1(\bm{r}|0)
\label{Abrikosov1}
\end{equation}
in the field range (\ref{secondllregion}), where $\varphi_1(\bm{r}|0) = {\hat a}^\dagger \varphi_0(\bm{r}|0)$, and $\varphi_0(\bm{r}|0)$ is given by (\ref{Abrikosov}). 
By minimizing (\ref{GL1}) with $\Delta=\Delta_1$ with respect to $A_1$ under fixed values of $k$ and $R$, we have the relation 
\begin{equation}
- \varepsilon_{h,1} + |A_1|^2 g \beta_A^{(1)} = 0, 
\label{MFEQ}
\end{equation}
where 
\begin{equation}
\varepsilon_{h,1} = \varepsilon_0 + 3 |c_1| h - 9 |c_2| h^2, 
\label{masssecond}
\end{equation}
and 
\begin{equation}
g \beta_A^{(1)} = h(g_1 + 4 g_2) \langle 0,0|0,0 \rangle + \biggl[ g 
+ h \left( \frac{1}{2} g_1 - 4 g_2 \right) \biggr] \langle 1,1|1,1 \rangle. 
\label{secondbeta1}
\end{equation}
Note that the combination $3(|c_1|-3 h |c_2|)$ in (\ref{masssecond}) is positive in the range (\ref{secondllregion}). In obtaining eq.(\ref{secondbeta1}), the second relation of eq.(\ref{bra1}) in Appendix was used. As far as $g \beta_A^{(1)}$ is positive, the stable 2LL vortex lattice structure is given by the function $\varphi_1(\bm{r}|0) = {\hat a}^\dagger \varphi_0(\bm{r}|0)$ with the parameters $k$ and $R$ minimizing $g \beta_A^{(1)}$. Since not only $\langle 1,1|1,1 \rangle$ but $\langle 0,0|0,0 \rangle$ affects $\beta_A^{(1)}$ in a manner dependent on the field $h$, the lattice structure, strictly speaking, depends on the ratios of $g_1 h/g$ and $g_2 h/g$, i.e., on the field value $h$. 
In fact, the lattice structures obtained in previous two works \cite{Klein,YM} are different in their orientations from each other. This fact can be understood by changing the parameters $g$, $g_1 h$, and $g_2 h$ within the present model \cite{YM}. As an example, we show the lattice structure following from the parameter values $g > 0$, $g_2 h = 0.15 g$, and $g_1 + 4 g_2=0$ in Fig.2 which is similar to that in Ref.\cite{YM}. This structure is a result of the $k$ and $R$ values minimizing the 
bracket $\langle 1,1|1,1 \rangle$, which is expressed as 
\begin{equation}
\langle 1,1|1,1 \rangle = \frac{k}{\sqrt{2 \pi}} \sum_{n,m} \biggl[ \frac{3}{4} - \frac{k^2}{2}(n^2+m^2) + \frac{k^4}{4}(n^2-m^2)^2 \biggr] e^{-k^2(n^2+m^2)/2} {\rm cos}(2 \pi R n m). 
\label{beta11}
\end{equation}
On the other hand, as is seen in eq.(\ref{secondbeta1}), inclusion of a contribution accompanied by the bracket $\langle 0,0|0,0 \rangle$ in the quartic term of the GL free energy tends to change the orientation of the structure. Figure 3 is the lattice structure following from the use of the parameter values $g > 0$, $g_1 h/g = - 0.5$, and $g_2 h/g = 0.15$. According to the results obtained in terms of different sets of the parameters, it appears that, if the coefficient $g_1+4 g_2$ of $\langle 0,0|0,0 \rangle$ in eq.(\ref{secondbeta1}) is positive as well as that of $\langle 1,1|1,1 \rangle$ there, the orientation of the structure of Fig.3 tends to be realized. In addition, if the coefficient of $\langle 1,1|1,1 \rangle$ is much smaller than that of $\langle 0,0|0,0 \rangle$ in eq.(\ref{secondbeta1}), the resulting structure becomes a nearly square one which is essentially different from the square lattice in 1LL \cite{AAA} (see the next paragraph). In this manner, since 2LL vortex lattice structure may be smoothly changed with increasing the magnetic field with no phase transition accompanied, one cannot regard Fig.2 simply as a rotated one of the structure of Fig.3. Such a field-dependent change of vortex lattice structures including the antivortices occurring with no transition accompanied has also been seen in the resulting phase diagrams of Rashba noncentrosymmetric superconductors \cite{Hiasa,Dan}. 

When $\varphi_1({\bf r}|{\bf r}_0)$ is expressed as $e^{-{\overline y}^2/2} \sigma_1(x,y)$ like eq.(\ref{Abrikosov}), $\sigma_1(x,y)$ depends not only on $\zeta = {\overline x}+i{\overline y}$ but also on $\zeta^*$. It implies that any 2LL vortex lattice may include {\it antivortices}. In fact, the presence of antivortices can be seen in the right figures of Fig.2 and 3 showing the spatial distributions of the gradient of the gauge-invariant phase, $\nabla \phi + 2 \pi {\bf A}/\phi_0$, for each case, where $\phi$ is the phase of $\Delta_1$. On the dark stripes in the left figures of Fig.2 and 3 showing the spatial variation of the amplitude $|\Delta_1|$ \cite{Klein,YM}, $|\Delta_1|$ is quite small but stays nonzero. The right figure in each case suggests that each dark stripe consists of a chain of the field-induced vortices and {\it antivortices}. Further, the nearly square lattice to be realized in 2LL mentioned in the preceding paragraph is also found to include the antivortices. This is in contrast to the square lattice in 1LL \cite{AAA} explained in sec.II which consists of the field-induced vortices with no antivortices. 

\begin{widetext}
\begin{figure}[t]
\scalebox{0.7}[0.7]{\includegraphics{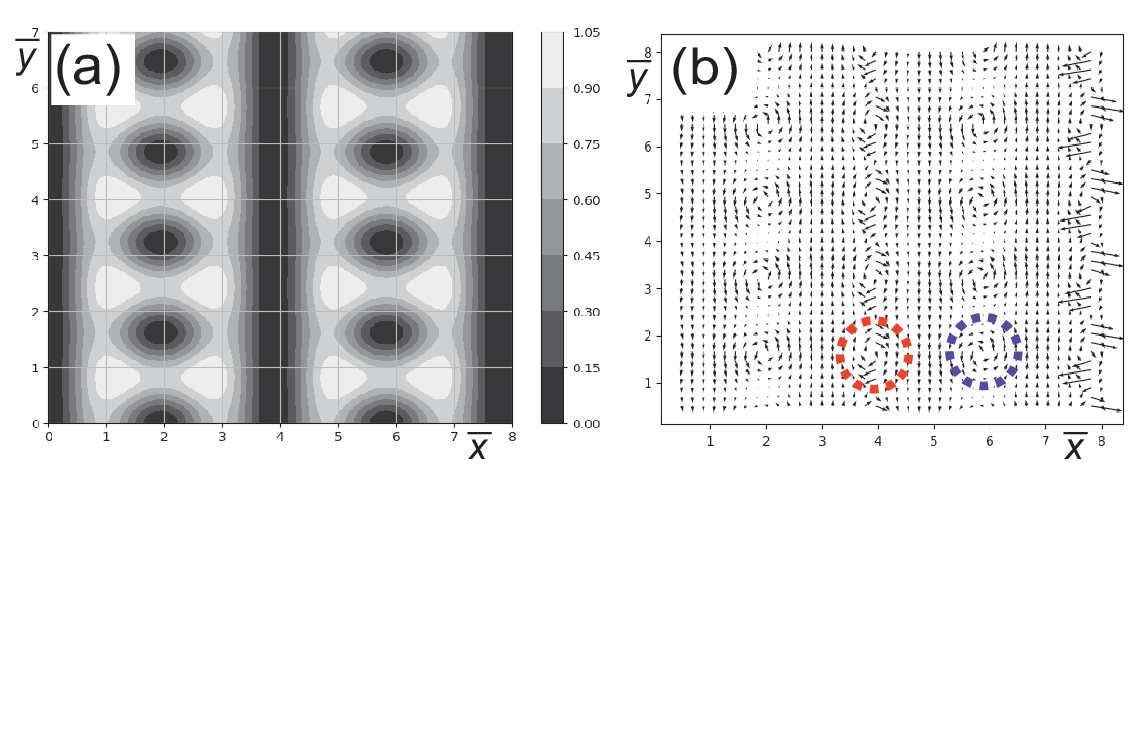}}
\caption{Spatial variations of (a) the amplitude and (b) the gauge-invariant gradient of the phase obtained in terms of the parameters $g_1 h/g = - 0.5$ and $g_2 h/g = 0.15$ with $g > 0$. The resulting parameters determining 2LL lattice structure become $k=1.577$ and $R=0$. As in Fig.2, the antivortices (red dashed circle) are clearly seen. }
\label{fig.3}
\end{figure}
\end{widetext}

In general, the form of the GL quartic term depends on the underlying electronic model. As far as the gradient expansion of the type done in eq.(\ref{GL1}) is useful, the model dependence should appear in relative differences between the coefficients $g$, $g_1$, and $g_2$. Within the weak-coupling approximation, a closed form of the quartic term can be obtained with no recourse to the gradient expansion \cite{AI03}, and its expression is shown in eqs.(\ref{QFgeneral}) and (\ref{betaAgeneral}) in Appendix. In the limit of small $h$ (see Appendix), eq.(\ref{betaAgeneral}) formally reduces to $\langle 1,1|1,1 \rangle$ given in eq.(\ref{beta11}), i.e., to the case of Fig.2. On the other hand, it is found that the vortex lattice structure following from the use of the general expression eq.(\ref{betaAgeneral}) is similar to Fig.3 \cite{Nunchot}. 

So far, we have focused on the region in the close vicinity of the $H_{c2}$-line so that the vortex solid below $H_{c2}$ at low temperatures is a purely 2LL state. Far from the $H_{c2}$-line, however, corrections from other LLs to the 2LL state are necessary to be incorporated. This may be performed based on eq.(\ref{MFEQ}) and by identifying an appropriate distance in the $H$-$T$ phase diagram from the $H_{c2}$-curve : In the case of the ordinary Abrikosov lattice described mainly by 1LL, the mean field solution can be generalized by expanding in powers of the distance from the $H_{c2}$-line, $\varepsilon_{h,0}/r_{10} = (H_{c2}(T) - B)/B$ \cite{Lasher,RI90}. Analogously, the corrections to the mean field solution in 2LL due to other LLs can be introduced in terms of the power expansion on the distances from the $H_{c2}$-curve, $\varepsilon_{h,1}/r_{01}$ and $\varepsilon_{h,1}/r_{21}$. 

Stability of 2LL vortex lattice is verified in a similar manner to the situation in 1LL case in sec.II. Consistently with the fact that the mean field solution is in 2LL of the SC order parameter, the intra LL 
fluctuation 
\begin{equation}
\delta \Delta_1 = A_1 [ a_{+1} \varphi_1(\bm{r}|\bm{r}_0) +  a_{-1} \varphi_1(\bm{r}|- \bm{r}_0) ]
\end{equation}
will be considered, and this and the mean field solution $\Delta_{1}$ will be substituted in ${\cal H}_{\Delta,1}$ and ${\cal H}_{\Delta,2}$ of (\ref{harmgl}) in Appendix as $\delta \Delta$ and $\Delta_{\rm MF}$ there, respectively. Then, the following two eigen values 
\begin{equation}
\xi_{\pm,1}({\bf r}_0) \equiv \frac{\varepsilon_{\pm,1}}{g |A_1|^4} = 2 \xi_{d,1}({\bf r}_0) - \xi_1 \pm |\xi_{n,1}({\bf r}_0)|
\end{equation}
are obtained as the dispersion relations of the resulting two eigenmodes within 2LL, where 
\begin{eqnarray}
2 \xi_{d,1}({\bf r}_0) - \xi_1 &=& 2 \langle 1+,1|1+,1 \rangle - \langle 1,1|1,1 \rangle + \frac{g_1 h}{g} \biggl( \, 2 \langle 2+, 1|2+, 1 \rangle + \langle 0+, 1|0+, 1 \rangle + 2 {\rm Re}\langle 2+, 1|1+, 2 \rangle \nonumber \\
&+& {\rm Re}\langle 0+, 1|1+, 0 \rangle - 2 \langle 1, 2|1, 2 \rangle - \langle 1, 0|1, 0 \rangle \, \biggr) + 2 \sqrt{2} \frac{g_2 h}{g} \biggl[ (2 \langle 1+, 1|0+, 2 \rangle - \langle 1,1|2,0 \rangle) + {\rm c.c.} \biggr],  \nonumber \\
\xi_{n,1}({\bf r}_0) &=& \langle 1,1|1+,1- \rangle + \frac{g_1 h}{g} \biggl(2 \langle 1, 2|1+, 2- \rangle + \langle 1,0|1+, 0- \rangle \biggr) + 2\sqrt{2} \frac{g_2 h}{g} \biggl[ \langle 1,1|2+,0- \rangle + {\rm c.c.} \biggr]. 
\label{shear1}
\end{eqnarray}
The dispersion relation of the low energy mode corresponding mainly to the phase fluctuation is $\varepsilon_{-,1}$. 
By expanding $\varepsilon_{-,1}$ in powers of ${\bf r}_0$, the absence of O(${\bf r}_0^2$) terms is straightforwardly checked. Concretely, the expression of $|\xi_{n,1}({\bf r}_0)|$ becomes 
\begin{eqnarray}
\biggl[1 &+& h \biggl(\frac{g_1 - 8 g_2}{2g} \biggr) \biggr] \biggl( \langle 1,1|1,1 \rangle \biggl[ 1 - \frac{5}{2} {\bf q}^2 r_B^2 \biggr] - 4 \sqrt{6} \, {\rm Re}(\langle 3,2|2,1 \rangle q_+^2 r_B^2) \biggr) + \langle 0,0|0,0 \rangle \biggl[ \frac{(g_1+4 g_2)h}{g} + \biggl(1 - 6 \frac{g_2 h}{g} \biggr) {\bf q}^2 r_B^2 \biggr] \nonumber \\
&+& 2 \biggl(1 - 6 \frac{g_2 h}{g} \biggr) \, {\rm Re}(\langle 1,1|0,0 \rangle q_+^2 r_B^2) 
\end{eqnarray}
up to O(${\bf r}_0^2$). Since the structures of 2LL vortex lattice are merely two-fold symmetric, the asymmetric brackets $\langle 1,1|0,0 \rangle$ and $\langle 3,2|1,2 \rangle$ accompanying the $q_+^2$-dependences are nonzero. Nevertheless, $\varepsilon_{-,1}$ becomes zero up to O(${\bf r}_0^2$) irrespective of the details of the lattice structure. Therefore, the energy arising from in-plane distortions of $\Delta$ of the minus (massless) mode in any 2LL vortex solid also becomes a sum of terms proportional to $q_x^4$, $q_y^4$, and $q_x^2 q_y^2$. 

\begin{figure}[t]
\scalebox{0.7}[0.7]{\includegraphics{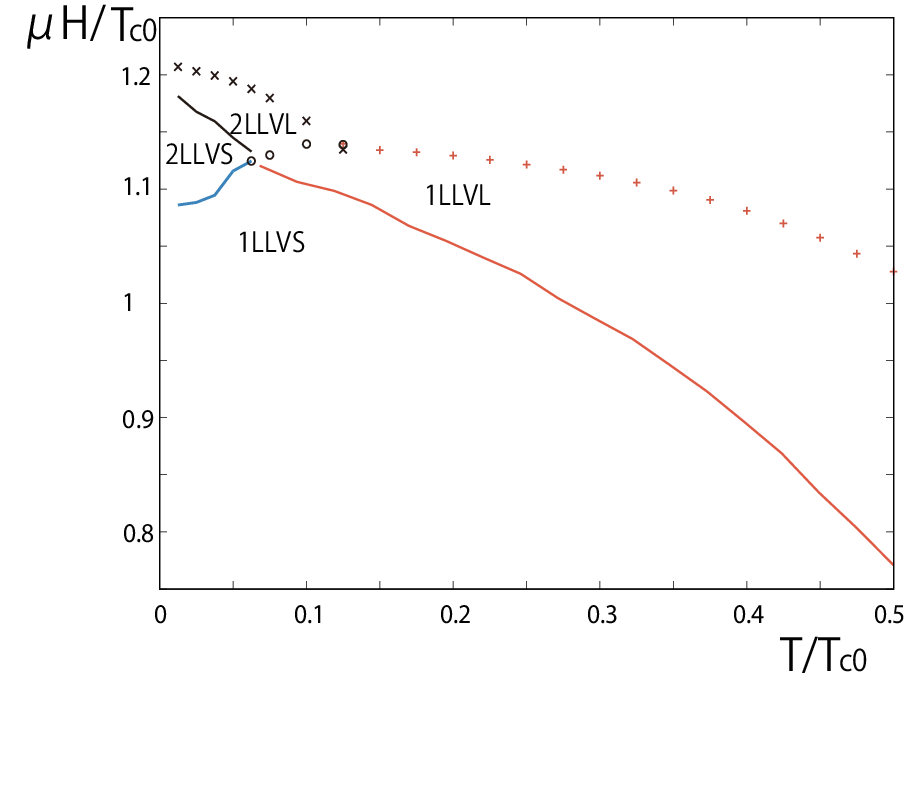}}
\caption{Example of the resulting phase diagram drawn in the case with the Maki parameter $\alpha_{\rm M} = 0.72$ by focusing on the high field and low temperature region. The ordinate is expressed by the Zeeman energy $\mu H$ divided by the zero field transition temperature $T_{c0}$. The cross (plus) symbols indicate the depairing line $H_{c2}(T)$ in 2LL (1LL) regime, and the red solid curve is the melting transition curve of the vortex solid formed in 1LL, 1LLVS, while the corresponding melting curve of the vortex solid in 2LL, 2LLVS, is the black solid one. The experimental irreversibility line \cite{Kasa20} consists of these two solid curves. Each of the remaining curve and/or symbols is explained in the main text. }
\label{fig.4}
\end{figure}
By analogy with 1LL case, this low energy mode corresponds to the incompressible elastic mode in 2LL vortex lattice because no gauge field fluctuation is excited through a coupling with the intra LL mode. Then, the vortex solid melting curve in 2LL regime may be examined based on the Lindemann criterion corresponding to eq.(\ref{Lindemann0}) \cite{Nakashima}. 

To estimate the melting line for 2LL vortex lattice based on the Lindemann criterion, the following simplification has been used: First, we focus on the case led to Fig.2 with $g > 0$, $g_1 h = -0.6 g$, and $g_2 h=0.15 g$, and the value of the coefficient $g$ will be identified with $7 \zeta(3) \beta^(t)_{A,1}/(8 \pi T)^2$ in the general expression (\ref{QFgeneral}) explained in Appendix in order to make the magnitude of the quartic term more realistic. Using the parameter values of $k$ and $R$ leading to Fig.2, $\xi_{-,1}({\bf r}_0)$ giving the dispersion relation of the minus mode in this case is found to be given by $0.308(q_x^4 + q_y^4)+0.908 q_x^2 q_y^2$. It corresponds to replacing $C_{66} {\bf q}_\perp^4$ in eq.(\ref{minusflucenergy}) by $2|A_1|^4 g (0.308(q_x^4 + q_y^4)+0.908 q_x^2 q_y^2)$. Further, by replacing $A_0$ in the first term of eq.(\ref{minusflucenergy}) and $a_{-0}$ in eq.(\ref{Lindemann0}) by $A_1$ and $a_{-1}$, respectively, an estimation of the melting transition line of 2LL vortex lattice can be obtained. 

In Fig.4, an example of the resulting $H$-$T$ phase diagram obtained in terms of the Maki parameter $\alpha_{\rm M} \equiv \mu H_{{\rm c2}, {\rm orb}}(0)/(2 \pi T_{c0}) =0.72$ is shown, where $\mu H$ is the Zeeman energy (see eq.(\ref{betaAgeneral})), and $H_{{\rm c2}, {\rm orb}}(T)$ is the $H_{c2}(T)$ in the absence of PPB. In Fig.4, the $H_{c2}(T)$-curve and the melting (black solid) curve in 2LL regime obtained according to the treatment explained above have been added in the $H$-$T$ phase diagram following from the approach developed in Ref. \cite{Nakashima} where no possibilities of 2LL vortex state have been taken into account. In fact, the 1LL melting (red solid) curve in Fig.4 is nearly the same as the blue curve in Fig.3 of Ref.\cite{Nakashima}. However, the transition curve to the 1LL FFLO state \cite{AI03} with a periodic modulation along the magnetic field direction in Fig.3 of Ref.\cite{Nakashima} is lost in the present Fig.4 due to the appearance of 2LL vortex solid (2LLVS). Note that, although, in the mean field approximation, the 1LL FFLO state should occur on the line indicated by the open circles, these symbols lie in either of the two vortex liquid regime, 1LLVL or 2LLVL, and thus, the resulting line consisting of the open circles cannot become a genuine phase transition one \cite{AI03}. Further, the first order transition (blue solid) curve between the 2LLVS and 1LLVS phases have been determined simply by comparing the condensation energy of 2LLVS with that of 1LLVS. We expect the novel SC phase detected in Ref.\cite{Kasa20} to correspond to 2LLVS surrounded by the black and blue solid curves in the figure. 

\section{Absence of vortex flow in 2LL vortex lattice}

As mentioned in the preceding section, the 2LL vortex lattices realized as an ordered state include antivortices. This fact has an important implication: Expected responses of a 2LL vortex lattice to an external electric current are sketched in the two figures of Fig.5. It is expected, as in the left figure of Fig.5, that, for a weak current parallel to the dark stripe with structure like a chain of the vortices and the antivortices, those dark stripes would not move while keeping the lattice structure entirely. It implies that the conventional uniform vortex flow does not occur in 2LL vortex lattices. Below, this conjecture will be tested by examining the superfluid stiffness or the helicity modulus. 

As in 1LL case, let us introduce an external disturbance of the gauge field  $\delta \bm{A}$ and the order parameter fluctuation $\delta \Delta$ coupling to $\delta \bm{A}$ in the general form of the harmonic fluctuation Hamiltonian (\ref{harmgl}) given in Appendix. In the case with the mean field solution $\Delta_{1} = A_1 \varphi_1(\bm{r}|0)$ belonging to 2LL, the order parameter fluctions $\delta \Delta$ coupling to $\delta \bm{A}$ are a linear combination of the functions in 1LL and the third ($n=2$) LL, i.e., $\varphi_0(\bm{r}|0)$ and 
$\varphi_2(\bm{r}|0)$. Hence, we will use 
\begin{equation}
\delta \Delta = A_1 [a_0 \, \varphi_0(\bm{r}|0) + a_2 \, \varphi_2(\bm{r}|0)]
\label{fluc2}
\end{equation}
below. By identifying (\ref{Abrikosov1}) and (\ref{fluc2}) with $\Delta$ and $\delta \Delta$ in eq.(\ref{harmgl}) in Appendix, respectively, and deleting $\varepsilon_0$ by using eq.(\ref{MFEQ}), we obtain the following ${\cal H}_{\Delta,1}+{\cal H}_{\Delta,2} = {\cal H}_{\Delta,s}+{\cal H}_{\Delta,as}$, where   
\begin{widetext}
\begin{eqnarray}
{\cal H}_{\Delta,s} &=& 2h|A_1|^2 \biggl( \biggl[ |c_1| - 4 |c_2| h \biggr]|a_0|^2 + \biggl[8 |c_2| h - |c_1| \biggr] |a_2|^2 \biggr) + g |A_1|^4 \biggl( \biggl[ 2 \langle 1,0|1,0 \rangle - \langle 1,1|1,1 \rangle \biggr] |a_0|^2 + \biggl[ 2 \langle 2,1|2,1 \rangle \nonumber \\
&-& \langle 1,1|1,1 \rangle \biggr] |a_2|^2 + \biggl[ \langle 1,1|2,0 \rangle a_0 a_2 + {\rm {c.c.}} \biggr] \biggr) + \frac{g_1 h}{2} |A_1|^4 \biggl( \biggl[ 2 \langle 2,0|2,0 \rangle - \langle 0,0|0,0 \rangle + \sqrt{2} (\langle 1,1|2,0 \rangle + {\rm {c.c.}}) \biggr]|a_0|^2 \nonumber \\
&+& \biggl[ 3 \langle 3,1|3,1 \rangle + 2 \langle 2,2|2,2 \rangle + \langle 2,0|2,0 \rangle + \langle 1,1|1,1 \rangle - 2 \langle 0,0|0,0 \rangle + \sqrt{2}(\sqrt{3} \langle 1,3|2,2 \rangle + \langle 1,1|2,0 \rangle + {\rm {c.c.}}) \biggr]|a_2|^2 \nonumber \\ 
&+& \frac{\sqrt{2}}{2} \biggl[ 2 \langle 1,0|1,0 \rangle + 2 \langle 2,1|2,1 \rangle + \sqrt{3}( \langle 2,1|3,0 \rangle + {\rm {c.c.}} ) \biggr] (a_0 a_2 + {\rm {c.c.}}) \biggr) + g_2 h |A_1|^4 \biggl( 4 \biggl[ \langle 1,0|1,0 \rangle + \langle 1,1|1,1 \rangle \nonumber \\ 
&-& \langle 0,0|0,0 \rangle \biggr]|a_0|^2 + 2 \biggl[ 4 \langle 2,1|2,1 \rangle + 2 (\langle 1,1|1,1 \rangle - \langle 0,0|0,0 \rangle) + \sqrt{3}(\langle 1,2|3,0 \rangle + {\rm {c.c.}}) \biggr]|a_2|^2 \nonumber \\
&+& \sqrt{2} \biggl[ \langle 1,1|1,1 \rangle 
+ 2 \langle 2,0|2,0 \rangle \biggr](a_0 a_2 + {\rm {c.c.}}) \biggr), 
\end{eqnarray}
and 
\begin{eqnarray}
{\cal H}_{\Delta, as} &=& |A_1|^4 \biggl(\frac{g}{2} \biggl[\langle 1,1|0,0 \rangle a_0^2 + \langle 2,2|1,1 \rangle (a_2^*)^2 + 4 \langle 1,0|2,1 \rangle a_0^* a_2 + {\rm {c.c.}} \biggr] + \frac{g_1}{2} h \biggl[ \sqrt{2} \biggl( \langle 1,2|1,0 \rangle a_0^2 + (\sqrt{3} \langle 3,2|2,1 \rangle \nonumber \\
&+& \langle 2,1|1,0 \rangle) (a_2^*)^2 \biggr) + \biggl( 2 \langle 2,0|2,2 \rangle + \langle 0,0|2,2 \rangle + \sqrt{3}(\langle 1,1|3,1 \rangle + \sqrt{2} \langle 2,0|3,1 \rangle) \nonumber \\ 
&+& \sqrt{2}(\langle 1,1|2,2 \rangle + \langle 0,0|1,1 \rangle) \biggr) a_0^* a_2 + {\rm {c.c.}} \biggr] + g_2 h \biggl[ \sqrt{2} \biggl( \langle 2,0|0,0 \rangle a_0^2 + (\langle 2,2|2,0 \rangle + \sqrt{3} \langle 3,1|1,1 \rangle)(a_2^*)^2 \biggr) \nonumber \\ 
&+& \sqrt{3} \biggl(\sqrt{3} \langle 1,0|2,1 \rangle + \langle 1,0|3,0 \rangle \biggr) a_0^* a_2 + {\rm {c.c.}} \biggr] \biggr).
\end{eqnarray}
\end{widetext}

\begin{figure}[t]
\scalebox{0.3}[0.3]{\includegraphics{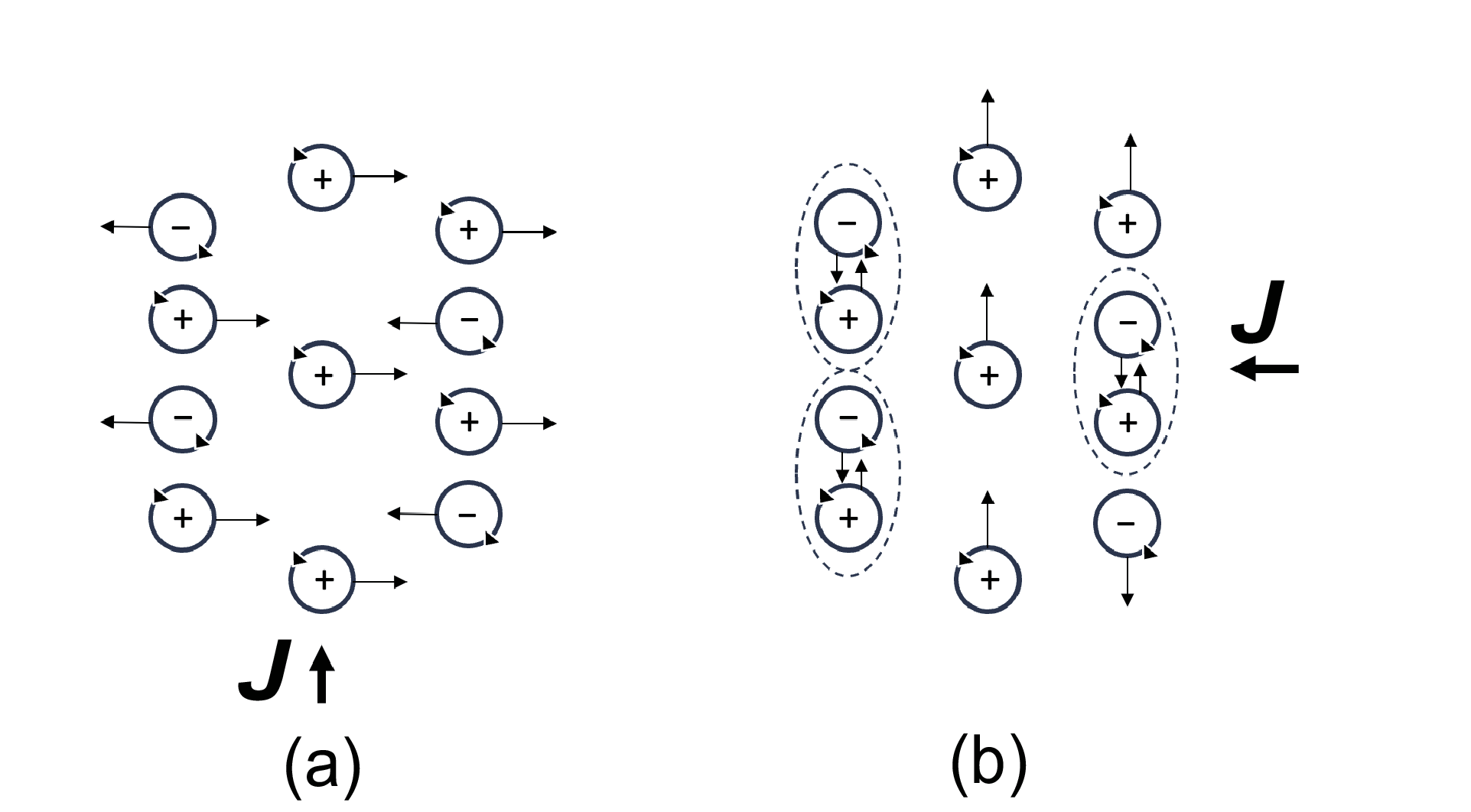}}
\caption{Schematic pictures of vortex motions expected in 2LL vortex solid of Fig.3 when a current $\bm{J}$ is applied (a) along the dark stripes (see Figs.2 and 3) of the lattice and (b) in the direction perpendicular to the stripes. The plus (minus) symbols indicate the field-induced vortices (antivortices). }
\label{fig.5}
\end{figure}

On the other hand, the contribution ${\cal H}_{A,1}$ proportional to $|A_1|^2$ and associated with $\delta \bm{A}$ is given by 
\begin{equation}
{\cal H}_{A,1} = |A_1|^2 \delta {\tilde A}_+ \delta {\tilde A}_- [12 |c_2| h - |c_1|] 
+ |A_1|^2 \biggl[ \frac{\delta {\tilde A}_+}{\sqrt{2h}} (- r_{01} a_0 + r_{21} \sqrt{2} a_2^*) + {\rm {c.c.}} \biggr]. 
\end{equation}
To express the remaining $\delta \bm{A}$-dependent contribution ${\cal H}_{A,2}$, it will be convenient to, as well as ${\cal H}_{\Delta,1}+{\cal H}_{\Delta,2}$, divide it into two contributions ${\cal H}_{A,2,s}$ expressed only by symmetric brackets and ${\cal H}_{A,2,as}$ which is nonvanishing when asymmetric brackets are nonzero. These two contributions are  
\begin{widetext}
\begin{eqnarray} 
{\cal H}_{A,2,s} &=& \left(\frac{g_1}{2} + g_2 \right)|A_1|^4 \biggl( \langle 1,1|1,1 \rangle \delta {\tilde A}_+ \delta {\tilde A}_- + \sqrt{\frac{h}{2}} \biggl(\delta {\tilde A}_+ \biggl[ (\langle 1,1|1,1 \rangle + 2 \langle 1,0|1,0 \rangle + \sqrt{2} \langle 1,1|2,0 \rangle) a_0 \nonumber \\ &+& (\sqrt{2}\langle 1,1|1,1 \rangle + \langle 2,0|1,1 \rangle + 2 \sqrt{2} \langle 2,1|2,1 \rangle) a_2^* \biggr] + {\rm {c.c.}} \biggr) \biggr), \nonumber \\
{\cal H}_{A,2,as} &=& \left(\frac{g_1}{2} + g_2 \right) \sqrt{\frac{h}{2}} \biggl(\delta {\tilde A}_+ \biggl[ (\langle 0,0|1,1 \rangle + 2 \sqrt{2} \langle 1,0|2,1 \rangle) a_0^* \nonumber \\ 
&+& (2 \langle 1,0|2,1 \rangle + \sqrt{2} \langle 1,1|2,2 \rangle + \sqrt{3} \langle 1,1|3,1 \rangle) a_2 \biggr] + {\rm {c.c.}} \biggr)
\end{eqnarray}
\end{widetext}

It is useful to arrange the expressions obtained above further. First, the sum ${\cal H}_{c_1,c_2}$ of ${\cal H}_{A,1}$ and the first two terms proportional to $|A_1|^2$ of ${\cal H}_{\Delta,s}$ simply becomes 
\begin{equation}
{\cal H}_{c_1,c_2} = |A_1|^2 ( r_{01} |{\tilde a}_0|^2 + r_{21} |{\tilde a}_2|^2). 
\label{hamil0}
\end{equation}
where 
\begin{eqnarray}
{\tilde a}_0 &=& a_0 - \frac{1}{\sqrt{2h}} \delta {\tilde A}_-, \nonumber \\
{\tilde a}_2 &=& a_2 + \frac{1}{\sqrt{h}} \delta {\tilde A}_+.
\end{eqnarray}
Next, the terms in ${\cal H}_{\Delta,s}$ proportional to the coefficient $g$ are summarized in the form 
\begin{equation}
{\cal H}_{g} = \frac{g}{2} |A_1|^4 \, (\langle 0,0|0,0 \rangle - \langle 1,1|1,1 \rangle) \, |\sqrt{2} {\tilde a}_0 + {\tilde a}_2^*|^2, 
\label{hamil1}
\end{equation}
and the sum of the remaining terms in ${\cal H}_{\Delta,s}$ and ${\cal H}_{A,2,s}$, denoted as ${\cal H}_{g_1,g_2,s}$ hereafter, becomes 
\begin{widetext}
\begin{eqnarray}
{\cal H}_{g_1,g_2,s} &=& h \left( \frac{g_1}{2} + g_2 \right)|A_1|^4 \biggl(-2 \langle 0,0|0,0 \rangle |{\tilde a}_0|^2 + (\langle 0,0|0,0 \rangle + \langle 1,1|1,1 \rangle)|{\tilde a}_2|^2 \biggr) + h |A_1|^4 \biggl(\frac{3}{4}g_1 \langle 0,0|0,0 \rangle \nonumber \\ 
&+& \biggl(-\frac{g_1}{4}+2g_2 \biggr) \langle 1,1|1,1 \rangle \biggr)|\sqrt{2} {\tilde a}_0 
+{\tilde a}_2^*|^2. 
\label{hamil2}
\end{eqnarray}
\end{widetext}
In obtaining the expressions arranged above, all of which are expressed only by ${\tilde a}_0$ and ${\tilde a}_2$, no knowledges on the vortex lattice structure were necessary. In fact, the relations shown in Appendix between various symmetric brackets are satisfied irrespective of the vortex lattice structure and result only from properties of the LLs. In contrast, the remaining terms proportional to the already defined asymmetric brackets are sensitive to the vortex lattice structure and, in general, are nonvanishing when the vortex lattice is less symmetric than the square lattice with the four-fold symmetry and the triangular lattice with the six-fold symmetry \cite{Lasher}. This asymmetric contribution arising from the sum of ${\cal H}_{\Delta,as}$ and ${\cal H}_{A,2,as}$ becomes 
\begin{widetext}
\begin{eqnarray}
{\cal H}_{g_1,g_2, as} &=& - \frac{\sqrt{2}}{4} |A_1|^4 \biggl[ \langle 20|00 \rangle \biggl( \, \biggl(g + \frac{3}{2}g_1 \biggr) (\sqrt{2} {\tilde a}_0 + {\tilde a}_2^*)^2 + h(g_1 + 2 g_2) (-2 {\tilde a}_0^2 + ({\tilde a}_2^*)^2 ) \biggr) + {\rm {c.c.}} \biggr] \nonumber \\ 
&+& \frac{\sqrt{2}}{4}|A_1|^4 \left[ (\langle 2,0|0,0 \rangle + \sqrt{2} \langle 2,2|1,1 \rangle) \biggl(g + h \biggl(\frac{g_1}{2} - 6 g_2 \biggr) \biggr) \biggl( {\tilde a}_2^* - \frac{A_1}{\sqrt{h}} \delta {\tilde A}_- \biggr)^2 + {\rm {c.c.}} 
\right].
\label{sllasymrosfinal}
\end{eqnarray}
\end{widetext}

The asymmetric bracket forms appearing in the expressions given above are explicitly given by 
\begin{eqnarray}
\langle 2,0|0,0 \rangle &=& \frac{k}{4 \sqrt{\pi}} \sum_{n,m} (-1 + k^2 (m-n)^2) e^{-k^2(n^2+m^2)/2} e^{-i 2 \pi R n m}, \nonumber \\ 
\langle 2,2|1,1 \rangle &=& \frac{k}{16 \sqrt{2 \pi}} \sum_{n,m} \biggl[ 3 - 3 k^2(2(n-m)^2 + (n+m)^2) + 4 \biggl(1 - \frac{k^2}{2}(n-m)^2 
\biggr)^2 - k^6(n-m)^2(n^2 - m^2)^2 \nonumber \\ 
&+& 6 k^4(n^2 - m^2)^2 \biggr] e^{-k^2(n^2+m^2)/2} 
e^{-i 2 \pi R n m}.
\end{eqnarray}
It is known that these asymmetric brackets vanish for the lattice structures such as those with the four-fold square symmetry or the six-fold hexagonal one \cite{Lasher}. In the case with merely a two-fold symmetry like Fig.1 and 2, however, these asymmetric brackets remain nonzero. As already suggested in eq.(\ref{flowLLL}), the superfluid stiffness defined by an electric current applied in a direction perpendicular to ${\bf B}$ becomes nonzero due to the presence of the remaining nonvanishing terms accompanied by asymmetric brackets. 

\begin{figure}
\scalebox{0.45}[0.45]{\includegraphics{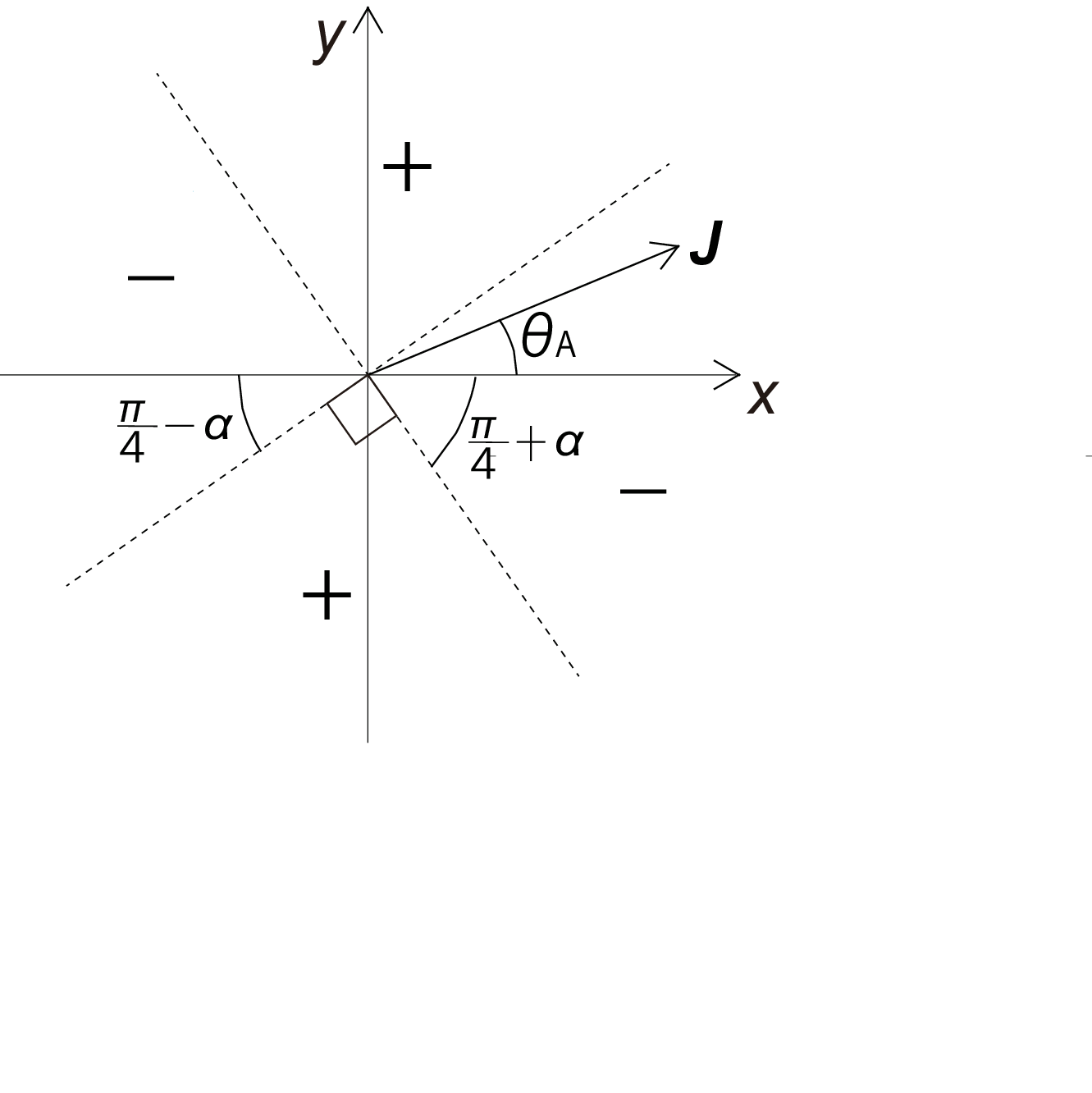}}
\caption{Relation between the sign of $\Upsilon_s$ and the direction $\bm{J}$ of an applied current in the plane perpendicular to the magnetic field in a 2LL vortex lattice. The angle $\alpha$ is the phase factor of the asymmetric brackets (see the main text). For instance, the conductivity diverges for a current applied in a direction satisfying $\pi/4 - \alpha < \theta_A < 3 \pi/4 - \alpha$. }
\label{fig.6}
\end{figure}

To see this concretely based on the expression (\ref{ros}), the $\delta \bm{A}$ dependences of the free energy $F(\delta \bm{A})$ in (\ref{ros}) will be examined. It is obtained by integrating over the fluctuation $\delta \Delta$ and keeping $\delta \bm{A}$ intact. In the case of the Meissner phase with no vortices, the transverse components of $\delta \bm{A}$ is not lost in the resulting $F(\delta \bm{A})$, and hence, $\Upsilon_s$ becomes a positive scalar which is nothing but the superfluid density. In the conventional triangular vortex lattice formed in 1LL of the SC order parameter, as seen in sec.III, the free energy corresponding to $F(\delta \bm{A})$ becomes independent of $\delta \bm{A}$ after integrating over the fluctuation $\delta \Delta_1$ in 2LL. In contrast, in the present case of 2LL vortex lattice, additional $\delta A_\pm$ dependence is present in the last terms of eq.(\ref{sllasymrosfinal}). The origin of the additional $\delta A_\pm$ dependence in eq.(\ref{sllasymrosfinal}) is the  spontaneously created anisotropy in the vortex structure which is at most two-fold symmetric. 

As already noted, the present analysis constructing the mean field vortex lattice by focusing only on 2LL is valid up to the lowest order in $\varepsilon_{h,1}/r_{01}$ and $\varepsilon_{h,1}/r_{21}$ which correspond to $[H_{c2}(T) - B]/B$ in the low field regime where the PPB is negligible. Then, as far as the lowest order results on $\Upsilon_s$ in $\varepsilon_{h,1}/r_{01}$ and $\varepsilon_{h,1}/r_{21}$ are concerned, one has only to set ${\tilde a}_0 = {\tilde a}_2 = 0$ in the sum of eqs.(\ref{hamil0}), (\ref{hamil1}), (\ref{hamil2}), and (\ref{sllasymrosfinal}) because the smallness of, for instance, $\varepsilon_{h,1}/r_{01}$ implies that the inequalities $r_{01} \gg {\rm Max}(|g \, A_1^2|, \,\,\,\,\,\, |g_1 h \, A_1^2|, ,\,\,\,\,\, |g_2 h \, A_1^2|)$ are satisfied. Then, we have 
\begin{equation}
\delta F(\bm{A}) = \sqrt{2} \frac{\pi}{\phi_0 B} |A_1|^4 \biggl(g + h \biggl(\frac{g_1}{2} - 6 g_2 \biggr) \biggr) |\langle 2,0|0,0 \rangle + \sqrt{2} \langle 2,2|1,1 \rangle| |\delta A_\perp|^2 \, {\rm cos}[2(\theta_A + \alpha)],   
\end{equation}
where the angles $\theta_A$ and $\alpha$ are defined in the manner 
\begin{eqnarray}
\langle 2,0|0,0 \rangle + \sqrt{2} \langle 2,2|1,1 \rangle &=& |\langle 2,0|0,0 \rangle + \sqrt{2} \langle 2,2|1,1 \rangle| e^{-i 2 \alpha}, \nonumber \\
\delta A_\pm &=& \delta A_\perp e^{\pm i \theta_A} 
\label{delF}
\end{eqnarray}
(see also Fig.6), and the $x$ and $y$ directions are defined according to the orientation of the lattice structure used in Figs.1, 2, and 3. 

This expression will be applied to the situations in which the structures of Fig.2 and 3 are realized. In the case of Fig.2, we have $\langle 2,0|0,0 \rangle + \sqrt{2} \langle 2,2|1,1 \rangle = 0.001914 - i 0.001709 \simeq 0.0027 e^{-i \pi/4}$ so that $\alpha \simeq \pi/8$, while $\langle 2,0|0,0 \rangle + \sqrt{2} \langle 2,2|1,1 \rangle = 0.1553$, i.e., $\alpha=0$ for the structure of Fig.3. Further, according to the parameter values of $g$, $g_1 h$, and $g_2 h$ used to obtain Fig.2 and 3 (see their captions), the coefficient $g+h(g_1/2 - 6 g_2)$ in eq.(\ref{delF}) is negative in both the cases of Fig.2 and 3. Then, the resulting superfluid stiffness $\Upsilon_s$ is nonzero in general and is found to have the following signs dependent on the direction of the current in each case: As sketched in Fig.6, $\Upsilon_s > 0$ in $\pi/4 < \theta_A + \alpha < 3 \pi/4$, while $\Upsilon_s < 0$ in $-\pi/4 < \theta_A + \alpha < \pi/4$. 
In the current directions with $\Upsilon_s > 0$, the 2LL vortex lattice does not move, resulting in the superconducting response and hence, the vanishing resistivity. On the other hand, in the directions with $\Upsilon_s < 0$, current-induced vortex motion would lead to some instability of the vortex structure. This correlation between the current or the eletric field direction and the sign of $\Upsilon_s$ leads to the scenario suggested in Fig.5. 

However, we cannot conclude at this stage that this consistency between the obtained sign of $\Upsilon_s$ and the picture in Fig.5 is generally satisfied, because, as already mentioned, our results are valid only up to the lowest order in $\varepsilon_{h,1}/r_{01}$ and $\varepsilon_{h,1}/r_{21}$. Nevertheless, since these expansion parameters depend on the temperature and the field $h$, it is not possible that, far from the $H_{c2}(T)$-line, $\Upsilon_s$ for the gauge disturbance in the plane perpendicular to the magnetic field in 2LL vortex solid becomes identically zero. Thus, the picture in Fig.6 that the sign of $\Upsilon_s$ in 2LL vortex solid depends on the relative direction between the applied current and the lattice structure should be generally valid. 

\section{Conclusion and Discussion}

Taking over the conjecture \cite{NNI} that the novel SC phase in the high field and low temperatures of FeSe should be a vortex solid or glass formed in the second Landau level (2LL) of the SC order parameter, we have investigated two characteristic features of the 2LL vortex solid in clean limit in the manner of comparing with those of the conventional 1LL vortex solid. First, it has been pointed out that the structure of the 2LL vortex solid is supported by incorporating antivortices in addition to the vortices compatible with the direction of the applied magnetic field. Next, as one of the equilibrium properties of this ordered state, the low energy mode corresponding to the elastic mode of the 2LL vortex solid has been examined. Using an approximately estimated dispersion relation of the low energy mode, the temperature at which the melting of the 2LL vortex solid occurs at each field has been examined, since the genuine SC phase transition of a 3D vortex solid in clean limit is believed to be the vortex solid-liquid transition. Further, the electromagnetic response of the 2LL vortex solid in clean limit has been studied by calculating the superfluid stiffness because the conventional vortex flow is not expected in the 2L lattice with the antivortices included. It has been shown that the superfluid stiffness becomes nonzero in general, and that its sign is reversed as the relative direction between the applied current and the vortex lattice structure is changed. It implies that, depending on the current direction, the resistivity becomes zero even with no vortex pinning, or that some type of the paramagnetic Meissner effect \cite{pararos1,pararos2,pararos3} may occur. 

It is unclear what the above-mentioned paramagnetic Meissner response occurring in the vortex solid in clean limit implies. In Fig.5, it was identified with an instability between the vortices and antivortices on the dark stripes in Fig.2 and 3. In real systems with pinning sites for the vortices and antivortices, however, any vortex (and antivortex) motion originating from the negative $\Upsilon_s$ would compete with the pinning effects. Further discussion in the situations with pinning effects will not be given here because the resulting phenomenon should be sample dependent, and detailed numerical studies would be needed. Nevertheless, it will be natural to expect that the above-mentioned competition between the vortex motion originating from the negative $\Upsilon_s$ and the pinning effect lead to an anomalously weak pinning effect at macroscopic scales.

The present work was supported by a Grant-in-Aid for Scientific Research [No.21K03468] from the Japan Society for the Promotion of Science.

\vspace{10mm}

\appendix{\bf {Appendix A}}

In this Appendix, various relations associated with the Landau level eigenfunctions are derived, and the general form of the harmonic fluctuation contributions to the GL Hamiltonian is presented. 

Throughout the manuscript, the applied magnetic field $\bm{B}$ is assumed to be directed to the $-z$-axis. As a gauge satisfying $\bm{\nabla} \times \bm{A} = - B {\hat e}_z$, we use $\bm{A}(\bm{r}) = B y {\hat e}_x$. Then, the 1LL basis functions consist of $\varphi_0(\bm{r}|\bm{r}_0) = e^{i y_0 x} \varphi_0(\bm{r}+\bm{r}_0|0)$ \cite{Eilenberger}, where 
\begin{equation}
\varphi_0(\bm{r}|0) = \left(\frac{k^2}{\pi} \right)^{1/4} \sum_n \, e^{i \pi R n^2} \, e^{i k n {\overline x} - \frac{1}{2} ({\overline y} + k n)^2}. 
\end{equation}
The corresponding basis functions in the ($n+1$)-th Landau level are created in the manner 
\begin{equation}
\varphi_n(\bm{r}|\bm{r}_0) = \frac{({\hat a}^\dagger)^n}{\sqrt{n!}} \varphi_0(\bm{r}|\bm{r}_0), 
\end{equation}
where 
\begin{equation}
{\hat a}^\dagger = \frac{1}{\sqrt{2}} \left( - i \frac{\partial}{\partial {\overline x}} + {\overline y} - \frac{\partial}{\partial {\overline y}} \right).  
\label{adagger}
\end{equation}
The corresponding lowering operator satisfying ${\hat a} {\hat a}^\dagger - {\hat a}^\dagger {\hat a} = 1$ is 
\begin{equation}
{\hat a} = \frac{1}{\sqrt{2}} \left( - i \frac{\partial}{\partial {\overline x}} + {\overline y} + \frac{\partial}{\partial {\overline y}} \right), 
\end{equation}
where ${\overline {\bm{r}}} = \bm{r}/r_B$, and 
$r_B = \sqrt{\phi_0/(2 \pi B)}$. 

According to Lasher \cite{Lasher}, the following relations 
\begin{eqnarray}
\varphi_0^*(\bm{r}|0) \, \varphi_1(\bm{r}|0) &=& - ({\hat a} \, \varphi_0(\bm{r}|0))^* \varphi_0(\bm{r}|0) + \varphi_0^*(\bm{r}|0) \, {\hat a}^\dagger \varphi_0(\bm{r}|0) \nonumber \\
&=& \frac{1}{\sqrt{2}} {\hat D}_+ |\varphi_0(\bm{r}|0)|^2
\end{eqnarray}
will be used hereafter, 
where 
\begin{equation}
{\hat D}_\pm = - i \frac{\partial}{\partial {\overline x}} \mp \frac{\partial}{\partial {\overline y}}.
\end{equation}
By repeating similar operations, we have 
\begin{eqnarray}
|\varphi_1(\bm{r}|0)|^2 &=& \left( 1 - \frac{1}{2} {\hat D}_+ {\hat D}_- \right) |\varphi_0(\bm{r}|0)|^2, \nonumber \\
\varphi_1^*(\bm{r}|0) \varphi_2(\bm{r}|0) &=& {\hat D}_+ \left( 1 - \frac{1}{4} {\hat D}_+ {\hat D}_- \right) |\varphi_0(\bm{r}|0)|^2, \nonumber \\ 
|\varphi_2(\bm{r}|0)|^2 &=& - \left( 1 - \frac{1}{8} {\hat D}_+^2 {\hat D}_-^2 \right) |\varphi_0(\bm{r}|0)|^2 + 2 |\varphi_1(\bm{r}|0)|^2,
\end{eqnarray}

To proceed further, the Fourier representation of the bilinear form of $\varphi_n(\bm{r}|0)$ will be used. 
For a general vortex lattice solution, $|\varphi_n(\bm{r}|0)|^2$ is periodic in the $x$-$y$ plane and has a Fourier representation using reciprocal lattice vectors $\bm{G}$. In particular, $|\varphi_0(\bm{r}|0)|^2$ will be expressed in the form 
\begin{equation}
|\varphi_0(\bm{r}|0)|^2 = \sum_{\bm{G}} F_{\bm{G}} e^{i \bm{G}\cdot\bm{r}}
\end{equation}
with $F_{\bm{0}}=1$ because of the normalization $\langle |\varphi_n(\bm{r}|0)|^2 \rangle_s = 1$. 
Hereafter, the notation 
\begin{equation}
\langle f(\bm{G}) \rangle_R = \sum_{\bm{G}} F_{-\bm{G}} f(\bm{G}) F_{\bm{G}}. 
\end{equation}
In the nonlinear terms in the GL free energy, we encounter the bracket form 
\begin{equation}
\langle n,m|p,s \rangle = \langle \varphi_n^*(\bm{r}|0) \varphi_m^*(\bm{r}|0) \varphi_p(\bm{r}|0) \varphi_s(\bm{r}|0) \rangle_s, 
\end{equation}
where $\langle \,\,\, \rangle_s$ denotes the space average. Its simplest one is the so-called Abrikosov factor appearing in representing the triangular vortex lattice in 1LL ($n=0$ LL), i.e., $\beta_A = \langle 0,0|0,0 \rangle = \langle 1 \rangle_R$. As another simple bracket form, let us rewrite $\langle 1,0|1,0 \rangle$, which can be expressed as 
\begin{eqnarray}
\langle 1,0|1,0 \rangle &=& \langle |\varphi_1(\bm{r}|0)|^2 |\varphi_0(\bm{r}|0)|^2 \rangle_s = \frac{1}{2} \langle |\varphi_0(\bm{r}|0)|^2 [(2 - {\hat D}_+ {\hat D}_-)|\varphi_0(\bm{r}|0)|^2] \rangle_s \nonumber \\ 
&=& \langle [\varphi_0^*(\bm{r}|0) \varphi_1(\bm{r}|0)]^* \varphi_0^*(\bm{r}|0) \varphi_1(\bm{r}|0) \rangle_s \nonumber \\ 
&=& \frac{1}{2} \langle |\varphi_0(\bm{r}|0)|^2 {\hat D}_+ {\hat D}_- |\varphi_0(\bm{r}|0)|^2 \rangle_s. 
\end{eqnarray}
By comparing the Fourier transforms of these expressions with each other, we find the relation 
\begin{equation}
\langle 1 \rangle_R = \langle \bm{G}^2 \rangle_R, 
\end{equation}
and hence, we
have 
\begin{equation}
\langle 1,0|1,0 \rangle = \frac{1}{2} \langle 0,0|0,0 \rangle = \frac{1}{2} \langle 1 \rangle_R, 
\label{identity0}
\end{equation}
and 
\begin{equation}
\langle 1,1|1,1 \rangle = \frac{1}{4} \langle \bm{G}^4 \rangle_R. 
\end{equation}
Similarly, the following relations between the brackets 
\begin{eqnarray}
\langle 2,0|2,0 \rangle &=& \frac{1}{2} \langle 1,1|1,1 \rangle, \nonumber \\
\langle 2,0|1,1 \rangle &=& \frac{2}{3}\sqrt{6} \langle 2,1|3,0 \rangle = \frac{\sqrt{2}}{2} (\langle 0,0|0,0 \rangle - \langle 1,1|1,1 \rangle), \nonumber \\ 
\langle 2,1|2,1 \rangle &=& \frac{1}{4} (\langle 0,0|0,0 \rangle + \langle 1,1|1,1 \rangle), \nonumber \\ 
\langle 2,2|2,2 \rangle &=& 2(\langle 0,0|0,0 \rangle - 2 \langle 1,1|1,1 \rangle) + \frac{1}{64} \langle \bm{G}^8 \rangle_R, \nonumber \\
\langle 3,1|3,1 \rangle &=& \frac{3}{2} (\langle 0,0|0,0 \rangle - 2 \langle 1,1|1,1 \rangle) + \frac{1}{96} \langle \bm{G}^8 \rangle_R, \nonumber \\
\langle 2,2|3,1 \rangle &=& \frac{\sqrt{6}}{2} (- \langle 0,0|0,0 \rangle + 3 \langle 1,1|1,1 \rangle - \frac{1}{96} \langle \bm{G}^8 \rangle_R), 
\label{bra1}
\end{eqnarray}
are additionally obtained, 
where the relation 
\begin{equation}
\langle \bm{G}^6 \rangle_R = 3 (3 \langle \bm{G}^4 \rangle_R - 4 \langle 1 \rangle_R),
\label{identity1}
\end{equation}. 
which is found by rewriting $\langle 2,1|2,1 \rangle$ in two different forms has been used. 

It should be stressed that these relations on the symmetric brackets of the form $\langle n,m|p,s \rangle$ with $n+m=p+s$ follow only from the relations on the LLs and are satisfied irrespective of 
the details of the vortex lattice structures. 

Similarly, the corresonding relations among the asymmetric brackets of the type $\langle n,m|p,s \rangle$ with $n+m=p+s \pm 2$ are also obtained using the Fourier transformed representation, and we have 
\begin{eqnarray}
\langle 0,0|1,1 \rangle &=& - \sqrt{2} \langle 0,0|2,0 \rangle = - \frac{1}{2} \langle \bm{G}_+^2 \rangle_R, \nonumber \\
\langle 1,0|2,1 \rangle &=& - \frac{1}{2} \langle 0,0|2,0 \rangle, 
\nonumber \\ 
\langle 1,0|3,0 \rangle &=& \frac{\sqrt{3}}{2} \langle 0,0|2,0 \rangle, 
\nonumber \\
\langle 2,1|3,2 \rangle &=& - \frac{\sqrt{6}}{192} \langle G_+^2 \bm{G}^4 \rangle_R, \nonumber \\
\langle 1,1|2,2 \rangle &=& \sqrt{2} \langle 0,0|2,0 \rangle + 2\sqrt{6} \langle 2,1|3,2 \rangle, \nonumber \\
\langle 2,0|2,2 \rangle &=& -2 \langle 0,0|2,0 \rangle - 2 \sqrt{3} \langle 2,1|3,2 \rangle, \nonumber \\
\langle 1,1|3,1 \rangle &=& - \sqrt{3} \langle 0,0|2,0 \rangle - 4 \langle 2,1|3,2 \rangle, \nonumber \\
\langle 2,0|3,1 \rangle &=&  - \frac{1}{\sqrt{2}} \langle 1,1|3,1 \rangle, 
\label{bra2}
\end{eqnarray}
where $G_+ = G_x + i G_y$, and the fact that the identities 
\begin{eqnarray}
\langle G_+^2 \bm{G}^6 \rangle_R &-& 15 \langle G_+^2 \bm{G}^4 \rangle_R + 120 \langle G_+^2 \rangle_R =0, \nonumber \\
\langle G_+^2 \bm{G}^2 \rangle_R &=& 3 \langle G_+^2 \rangle_R 
\label{Fourierreciprocal}
\end{eqnarray}
similar to (\ref{identity0}) and (\ref{identity1}) are satisfied has been 
used. Each term of the first relation of eq.(\ref{Fourierreciprocal}) vanishes when the vortex lattice has square or hexaganal symmetry \cite{Lasher}. 

Next, the general form of the harmonic fluctuation contribution to the GL Hamiltonian (\ref{GL1}) will be given. By writing the order parameter field $\Delta$ as the sum of its mean field $\Delta_{\rm MF}$ and its fluctuation $\delta \Delta$, the harmonic fluctuation Hamiltonian takes the form ${\cal H}_{\Delta,1} + {\cal H}_{\Delta,2} + {\cal H}_{A,1} + {\cal H}_{A,2}$, 
where 
\begin{widetext}
\begin{eqnarray}
{\cal H}_{\Delta,1} &=& \int d^3\bm{r} \left[ - \varepsilon_0 |\delta \Delta|^2 - c_2 h^2 |(2{\hat a}^\dagger {\hat a} + 1)\delta \Delta|^2 + c_1 h \delta \Delta^* (2{\hat a}^\dagger {\hat a} + 1) \delta \Delta + \frac{g}{2} \biggl( 4 |\Delta_{\rm MF}|^2 |\delta \Delta|^2 + ( \, (\Delta_{\rm MF}^*)^2 (\delta \Delta)^2 + {\rm c.c.} \, ) \biggr) \right], \nonumber \\
{\cal H}_{\Delta,2} &=& \frac{g_1 \, h}{2} \int d^3\bm{r} \biggl[ |\delta \Delta|^2 (|{\hat a} \Delta_{\rm MF}|^2 + |{\hat a}^\dagger \Delta_{\rm MF}|^2) + |\Delta_{\rm MF}|^2 (|{\hat a} \delta \Delta|^2 + |{\hat a}^\dagger \delta \Delta|^2) 
+ \biggl( \, \Delta_{\rm MF}^* \delta \Delta \biggl[ ({\hat a}^\dagger \delta \Delta)^* {\hat a}^\dagger \Delta_{\rm MF} + ({\hat a} \delta \Delta)^* {\hat a} \Delta_{\rm MF} \nonumber \\ 
&+& ({\hat a}^\dagger \Delta_{\rm MF})^* {\hat a}^\dagger \delta \Delta 
+ ({\hat a} \Delta_{\rm MF})^* {\hat a} \delta \Delta \biggr] + {\rm {c.c.}} \biggr) \biggr] + g_2 \, h \int d^3\bm{r} \biggl[ \biggl( (\delta \Delta^*)^2 {\hat a}^\dagger \Delta_{\rm MF} {\hat a} \Delta_{\rm MF} + (\Delta_{\rm MF}^*)^2 {\hat a}^\dagger \delta \Delta {\hat a} \delta \Delta \nonumber \\
&+& 2(\Delta_{\rm MF} \delta \Delta)^* ({\hat a}^\dagger \Delta_{\rm MF} {\hat a} \delta \Delta + {\hat a}^\dagger \delta \Delta {\hat a} \Delta_{\rm MF} ) \biggr) + {\rm {c.c.}} \biggr], \nonumber \\
{\cal H}_{A,1} &=& \int d^3\bm{r} \biggl[ \delta{\tilde A}_+ \delta{\tilde A}_- \biggl( c_1 |\Delta_{\rm MF}|^2 - h c_2 [ \Delta_{\rm MF}^* (2 {\hat a}^\dagger {\hat a} + 1) \Delta_{\rm MF} + {\rm {c.c.}} ] \biggr) - 2 h c_2 |(\delta{\tilde A}_+ {\hat a} + \delta{\tilde A}_- {\hat a}^\dagger) \Delta_{\rm MF}|^2 \nonumber \\
&+& \sqrt{\frac{h}{2}} \biggl( \delta{\tilde A}_+ \biggl[ ({\hat a} \, \delta \Delta)^* (-2 c_2 h(2{\hat a}^\dagger {\hat a} + 1) + c_1) \Delta_{\rm MF} + ({\hat a} \, \Delta_{\rm MF})^* (-2 c_2 h(2{\hat a}^\dagger {\hat a} + 1) + c_1) 
\delta \Delta \nonumber \\
&+& [(-2 c_2 h(2{\hat a}^\dagger {\hat a} + 1) + c_1) \Delta_{\rm MF}]^* \, {\hat a}^\dagger \, \delta \Delta + [(-2 c_2 h(2{\hat a}^\dagger {\hat a} + 1) + c_1) \delta \Delta]^* \, {\hat a}^\dagger \Delta_{\rm MF} \biggr] + {\rm {c.c.}} \biggr) \biggr], \nonumber \\
{\cal H}_{A,2} &=& \int d^3\bm{r} \biggl[ \left( \frac{g_1}{2} + g_2 \right) \xi_0^2 |\Delta_{\rm MF}|^4 \delta{\tilde A}_+ \delta{\tilde A}_- + \frac{g_1}{2} \sqrt{\frac{h}{2}} \biggl( \delta{\tilde A}_+ \delta_1[ |\Delta|^2 ( \Delta^* ({\hat a}^\dagger \Delta) 
+ ({\hat a} \Delta)^* \Delta ) ] + {\rm {c.c.}} \biggr) \nonumber \\
&+& g_2 \sqrt{\frac{h}{2}} \biggl( \delta{\tilde A}_+ \delta_1[(\Delta({\hat a} \Delta))^* \Delta^2 
+ (\Delta^*)^2 (\Delta({\hat a}^\dagger \Delta))] + {\rm {c.c.}} \biggr) \biggr], 
\label{harmgl}
\end{eqnarray}
\end{widetext}
where $\delta{\tilde A}_\pm = 2 \pi \xi_0 (\delta A_x \pm i \delta A_y)/\phi_0$. Further, the symbol $\delta_1[f]$ implies that just one $\Delta$ (or $\Delta^*$) included in $f$ is replaced by $\delta \Delta$ (or $\delta \Delta^*$), while the remaining $\Delta$ and $\Delta^*$ are replaced by the mean field ones $\Delta_{\rm MF}$ and $\Delta_{\rm MF}^*$, respectively. 
In writing ${\cal H}_{\Delta,2}$, the relations
\begin{eqnarray}
(\bm{\Pi} A)^* \cdot (\bm{\Pi} B) &=& ({\hat a}^\dagger A)^* ({\hat a}^\dagger B) + ({\hat a} A)^* ({\hat a} B), \nonumber \\
(\bm{\Pi} A) \cdot (\bm{\Pi} B) &=& ({\hat a}^\dagger A) ({\hat a} B) + ({\hat a} A) ({\hat a}^\dagger B)
\end{eqnarray}
were used.

\vspace{10mm}

\appendix{\bf {Appendix B}}

Generally, the quartic (O($|\Delta|^4$)) term, $F^{(4)}_{\rm GL}$, of the SC contribution to the free energy is of a spatially nonlocal form when the applied magnetic field is present. In the state of the vortex lattice in 2LL where the order parameter $\Delta$ is expressed by $\Delta_1({\bf r}) = A_1 \varphi_1({\bf r}|0)$, the quartic term takes the form 
\begin{equation}
\frac{F^{(4)}_{{\rm GL},1}}{N(0) \Omega} = \frac{1}{2} \biggl\langle V_4({\bf r}_1, {\bf r}_2 ; {\bf r}_3, {\bf r}_4) \Delta_1^*({\bf r}_1) \Delta_1^*({\bf r}_2) \Delta_1({\bf r}_3) \Delta_1({\bf r}_4) \biggr\rangle_s \simeq \frac{7 \zeta(3)}{2 (8 \pi T)^2} \beta^(t)_{A,1} |A_1|^4.  
\label{QFgeneral}
\end{equation}
In the parameter representation used in Ref.\cite{AI03}, this factor $\beta^{(t)}_{A,1}$ is a function of the magnetic field and given by 
\begin{eqnarray}
\beta^{(t)}_{A,1} &=& \frac{(4 \pi t)^2}{7 \zeta(3)} \frac{k}{\sqrt{2 \pi}} \int_0^\infty \prod_{j=1}^3 d\rho_j \frac{2 \pi t \, {\rm cos}(2 \mu H \rho_{\rm tot})}{{\rm sinh}(2 \pi t \rho_{\rm tot})} \sum_{n,m} \exp\biggl(-\frac{k^2}{2}(n^2+m^2) \biggr) \, {\rm Re}\biggl( e^{i 2 \pi R n m} \biggl\langle |w_{{\hat {\bf p}}}|^4 \biggl(2(1-|\mu|^2 \rho_{\rm tot}^2) \nonumber \\ 
&+& \biggl[1 + \frac{(\mu^*)^2 \rho_{\rm tot}^2}{2} - [\sqrt{2} \rho_2 {\rm Re}\mu + k(n+m)]^2 \biggr] \biggl[ 1 + \frac{\mu^2 \rho_{\rm tot}^2}{2} - [\sqrt{2} (\rho_1 - \rho_3) {\rm Re}\mu + k(n-m)]^2 \biggr] \biggr) \nonumber \\
&\times& \exp\biggl(-\frac{|\mu|^2}{2}(\rho_2(\rho_1+\rho_3) + \sum_{j=1^3} \rho_j^2) - \frac{1}{4} (\mu^2 \rho_2^2 + (\mu^*)^2(\rho_1-\rho_3)^2) \nonumber \\ 
&+& \frac{k}{\sqrt{2}}(\mu^*(\rho_3-\rho_1)(n-m)-\mu \rho_2(n+m)) \biggr) \biggr\rangle_{{\hat {\bf p}}} \biggr)
\label{betaAgeneral}
\end{eqnarray}
in a magnetic field perpendicular to the basal plane of a material with a quasi 2D Fermi surface, where $\mu H$ is the Zeeman energy, $\rho_{\rm tot}=\sum_{j=1}^3 \rho_j$, $\mu = \sqrt{2 h} \pi ({\hat p}_x - i {\hat p}_y)$, and $t=T/T_{c0}$. 

\end{document}